\documentclass[twocolumn]{aastex61}
\usepackage{amsmath,amstext}
\usepackage[T1]{fontenc}
\usepackage{apjfonts} 
\usepackage[figure,figure*]{hypcap}
\usepackage{rotating, graphicx}
\usepackage{booktabs,threeparttable}
\usepackage{natbib}
\usepackage{geometry}
\usepackage{ulem}

\defcitealias{jennings2014}{J14}

\shorttitle{Supernova Progenitors in M83}
\shortauthors{Williams et al.}
\begin{document}

\title{The Masses of Supernova Remnant Progenitors in M83}

\author{Benjamin F. Williams}
\affiliation{Department of Astronomy, Box 351580, University of Washington, Seattle, WA 98195; ben@astro.washington.edu; tjhillis@uw.edu; 
rubab@uw.edu;
jd@astro.washington.edu}

\author{Tristan J. Hillis}
\affiliation{Department of Astronomy, Box 351580, University of Washington, Seattle, WA 98195; 
ben@astro.washington.edu; tjhillis@uw.edu; 
rubab@uw.edu;
jd@astro.washington.edu}

\author[0000-0003-2379-6518]{William P. Blair}
\affiliation{Department of Physics and Astronomy, Johns Hopkins University, 3400 North Charles Street, Baltimore, MD 21218; wblair@jhu.edu}

\author[0000-0002-4134-864X]{Knox S. Long}
\affil{Space Telescope Science Institute,
3700 San Martin Drive,
Baltimore MD 21218, USA; long@stsci.edu}
\affil{Eureka Scientific, Inc.
2452 Delmer Street, Suite 100,
Oakland, CA 94602-3017}

\author{Jeremiah W. Murphy}
\affiliation{Department of Physics, Florida State University, jwmurphy@fsu.edu}
\author{Andrew Dolphin}
\affiliation{Raytheon, 1151 E. Hermans Road, Tucson, AZ 85706; adolphin@raytheon.com}

\author{Rubab Khan}
\affiliation{Department of Astronomy, Box 351580, University of Washington, Seattle, WA 98195; 
ben@astro.washington.edu; tjhillis@uw.edu; 
rubab@uw.edu;
jd@astro.washington.edu}

\author{Julianne J. Dalcanton}
\affiliation{Department of Astronomy, Box 351580, University of Washington, Seattle, WA 98195; ben@astro.washington.edu; tjhillis@uw.edu; 
rubab@uw.edu;
jd@astro.washington.edu}

\begin{abstract}
We determine the ages of the young, resolved stellar populations at the locations of 237 optically-identified supernova remnants in M83. These age distributions put constraints on the progenitor masses of the supernovae that produced 199 of the remnants.  The other 38 show no evidence for having a young progenitor and are therefore good Type Ia SNR candidates.  Starting from Hubble Space Telescope broadband imaging, we measured resolved stellar photometry of seven archival WFC3/UVIS fields in F336W, F438W, and F814W.  We generate color-magnitude diagrams of the stars within 50~pc of each SNR and fit them with stellar evolution models to obtain the population ages. From these ages we infer the progenitor mass that corresponds to the lifetime of the most prominent age that is $<$50 Myr.  In this sample, there are 47 SNRs with best-fit progenitor masses $>$15~M$_{\odot}$, and 5 of these are $>$15~M$_{\odot}$ at 84\% confidence.  This is the largest collection of high-mass progenitors to date, including our highest-mass progenitor inference found so far, with a constraint of $<$8 Myr.  Overall, the distribution of progenitor masses has a power-law index of $-3.0^{+0.2}_{-0.7}$, steeper than Salpeter initial mass function ($-2.35$).  It remains unclear whether the reason for the low number of high-mass progenitors is due to the difficulty of finding and measuring such objects or because only a fraction of very massive stars produce supernovae.
\end{abstract}


\section{Introduction}

One of the most fundamental predictions from stellar evolution theory is that stars above a certain mass end their lives in powerful supernova (SN) explosions.  SNe play major roles in exciting and enriching the interstellar medium, as well as in regulating star formation and creating compact objects that can become significant emitters of X-rays, Gamma-rays, and even gravitational waves. 

Nearby SNe are well-observed and cataloged \citep[e.g.,][]{guillochon2017}.  However, they are limited in number and the masses of the progenitor stars are often very difficult to discern. Traditionally, direct imaging of progenitors has been the method of choice for constraining core-collapse supernovae (CCSNe) progenitor masses, but it has substantial limitations.  High-quality precursor imaging must exist, and the astrometry must be sufficiently accurate ($\sim$0.1$''$) to identify the exact star that exploded.   Furthermore, even when the progenitor is identified, interpretation of its photometry depends on the most uncertain stages of stellar evolution \citep{gallart05}.   The mass of a precursor is typically estimated by fitting stellar evolution models to the star's color and magnitude \citep[e.g.][]{chen2015}; however, mass loss, binary evolution, pulsation, internal mixing, and the formation of dust in stellar winds all contribute to systematic and random uncertainties in late-stage evolutionary models. Therefore, matching a stellar evolution model to a single evolved star places uncertain constraints on the initial mass. 

To date, archives of extragalactic resolved stellar imaging from the
{\it Hubble Space Telescope (HST)} have yielded 68 direct-image
constraints; of these, only 30 are detections, and the rest are upper
limits \citep[e.g.,][for recent reviews]{smartt2015,vandyk2017}.
While the statistics from the direct-imaging technique are low, they
already lead to some interesting conclusions and hints.  For example,
HST archival data shows that the most common supernovae (SNe), Type
II-P, map to red supergiant progenitors \citep{2009ARA&A..47...63S}.
\citet{smartt2015} present 18 detections and 27 upper limits and infer
a minimum mass of 7$^{+4}_{-1}$ M$_{\odot}$ for SN IIP.  They also
infer an upper mass for SN IIP of $\sim$17 M$_{\odot}$.  Yet, red
supergiant masses are estimated up to $\sim$25 M$_{\odot}$.  As a
result, they conclude that the most massive red supergiant progenitors
may not explode as SNe.  More recently, \citet{davies2017} suggest
that the bolometric corrections for red supergiants that are about to
explode are much larger than were used in the \citet{smartt2015}
analysis.  Using bolometric corrections calibrated with very
evolved red supergiants, \citet{davies2017} infer a maximum mass of
$27$ M$_{\odot}$, suggesting that even the most massive red
supergiants may indeed explode.  There are very few potential
  direct-imaging progenitor estimates for type Ib or Ic SNe, which
  have been difficult to interpret but may suggest high mass
  progenitors
  \citep[e.g.,][]{cao2013,groh2013,bersten2014,fremling2016,eldridge2016,vandyk2018,kilpatrick2018,xiang2019}.
Because of the small sample of direct-imaging measurements, the
mapping between progenitors and SN type remains unclear
\citep[e.g.][]{2009MNRAS.395.1409S,2012MNRAS.424.1372A,smartt2015,2017MNRAS.469.1445A,davies2017}.

 There are a few methods of indirectly constraining progenitor masses \citep[e.g.,][]{2012MNRAS.424.1372A}.  The technique we apply in this paper is age-dating of the surrounding stellar population of a SNR.  This method relies on the fact that only relatively massive stars ($>$7.5~M$_{\odot}$) become CCSNe \citep{jennings2012,diaz2018}.  Thus, their lifetimes are limited to $<$50~Myr \citep{girardi2002}. Over 90\% of stars form in clusters containing more than 100 members with $M_{\rm cluster}{>}50 {\rm M}_{\odot}$ \citep{lada03}.  These stars remain spatially correlated on physical scales up to $\sim$100~pc during the 100 Myr lifetimes of $4{\rm M}_{\odot}$ stars, even if the cluster is not gravitationally bound \citep{2006MNRAS.369L...9B}.  We have confirmed this expectation empirically in several cases \citep{badenes2009,2009ApJ...703..300G,2011ApJ...742L...4M,williams2018}.  Thus, it is reasonable to assume a physical association between any observed CCSN and other young stars of the same age and in the same vicinity.  We use the median star formation age, going to 50 Myr ($7.3 {\rm M}_{\odot}$ precursor equivalent, see Section~\ref{sec:progenitorMass} for details), to infer a progenitor mass using our chosen models.    Therefore, once the surrounding population age is constrained, one may infer the precursor mass through stellar evolution models without the need for precursor imaging (see \citealp{jennings2012}; \citealp{jennings2014} hereafter J14; \citealp{2014ApJ...791..105W}, for more details on the technique). 

Using this population age-dating method, one may estimate a progenitor mass for any location within 8 Mpc where there has been a CCSN.  To date, there are only 39 observed SNe within this volume \citep{guillochon2017}; however, there are thousands of supernova remnants (SNRs) in this volume.  These SNRs mark the locations of SNe for $\sim$20 kyr or more \citep{braun1989}, and if they reside in regions with recent star formation, they are likely to be from CCSNe.  Thus, if we use SNRs for measuring progenitor masses, we improve our time baseline for finding CCSNe locations by a factor of more than 200.

Studies of precursor masses of SNRs have yielded promising results on constraining the distribution of stellar masses that produce supernovae.  J14 constrained progenitor masses for 115 SNRs in M31 and M33, finding that their mass distribution was steeper than a standard Salpeter initial mass function (IMF), and detecting a clear lower mass limit to their distribution at ${\sim}7.5 {\rm M}_{\odot}$. \citet{diaz2018} updated and re-examined 94 of those SNRs using a more uniform photometric sample and a Bayesian treatment of contamination and uncertainties.  They confirm the J14 results, and place reliable uncertainties on both the power-law index of the mass distribution ($-3.0_{-0.3}^{+0.5}$) and the lower mass cutoff ($7.33_{-0.16}^{+0.02}$).  

Herein, we seek to extend these studies to the nearby galaxy M83.  M83 is a strong candidate for this kind of study because of its proximity \citep[4.61 Mpc;][]{saha2006}, a nearly face-on view at \citep[$i$=24$^o$;][]{lundgren2004}, and its long history of SNe \citep{richter1984}.  As searches for SNRs within M83 \citep{2004ApJS..155..101B,dopita2010,blair2012,2014ApJ...788...55B} have shown, this galaxy is rich with past supernova activity, with 307 SNRs and SNR candidates, which is more than the combined M31+M33 sample of J14.  This gives us improved statistical power, for the progenitor masses of SNe.  With such a large sample we can place improved constraints on the distribution of progenitor masses and on the upper mass limit for CCSNe progenitors. 

\indent In this paper, Section 2 discusses the archival {\it HST} data, and the analysis technique we use for estimating progenitor masses. Section 3 details our resulting progenitor age and mass estimates. Section 4 investigates the distribution of progenitor masses in the context of standard mass functions, and we conclude with a brief summary in Section 5. We assume a distance of 4.61 Mpc \citep{saha2006} throughout the paper, and we use the Padova \citep{girardi2002,2008A&A...482..883M, girardi2010} stellar evolution model library for all of the model fitting and for all stellar lifetime estimates.

\section{Data \& Analysis}

We analyze HST observations of seven WFC3/UVIS fields in three filters F336W, F438W, and F814W covering most of the high surface brightness portion of M83.  These observations are described in detail in \citet{2014ApJ...788...55B}. In the subsections that follow, we discuss the origin of the SNR positions we analyzed, our technique for measuring resolved stellar photometry, our method for fitting models to the photometry to determine the ages of the stellar populations, and our process for inferring progenitor masses from these ages. 
    
    \subsection{Supernova Remnant Locations}

We took the locations of SNRs and strong SNR candidates in M83 from a compilation of catalogs obtained by several surveys. We provide the full combined catalog in Table~\ref{tab:catalog}.  The survey where most of the SNRs were discovered was from the Magellan 6.5 m telescope at Las Campanas \citep[][B12 in the Table]{blair2012}.  Those SNR candidates found first using HST are from \citet{dopita2010} and \citet{2014ApJ...788...55B}, which are D10 and B14 in the Table. We note that the \citet{dopita2010} catalog covered the complex starburst nucleus where confusion effects prevent resolved photometry even for HST, so many of these objects are excluded from the photometric analysis.  There are a handful of SNR candidates in both the Magellan data set and the HST data that have not been reported previously but are included here for completeness.  For the SNRs that have identified X-ray counterparts from \citet[][L14 in the Table]{long14}, a cross reference is also provided. Spectroscopic confirmations of a significant subsample of these SNRs were reported by \citet{winkler2017}.   Furthermore, there are six historical SNe whose locations are well-determined, which we also include in the sample.  An additional young SNR that may represent an unobserved SN in the last century, SNR~238 (B12-174a), is also included \citep{blair2015}. Table\ref{tab:catalog} represents the most complete listing of SNRs and good SNR candidates in M83 to date. 

Table\ref{tab:catalog} also includes a quality flag for each SNR candidate, based on the amount of supporting evidence used to confirm the object is a SNR.  The quality flag values indicate the following categories: 
    
\noindent
1a:  Optical imaging candidate confirmed by spectroscopy and having X-ray emission.\\
1b:  Optical imaging candidate strongly confirmed by optical spectroscopy (e.g. spectroscopic [S~II]:H$\alpha$ > 0.5) but no X-ray detection.\\
1c:  Optical imaging candidate with photometric [S~II]:H$\alpha$ > 0.5 but no optical spectral confirmation available and no X-ray detection.\\
1d:  Optical imaging candidate with X-ray but no spectra available.\\
1e:  Optical imaging candidate with X-ray but marginal spectral ratio (likely due to H~II contamination).\\
1f:  Solid Imaging candidate, no X-ray, but with clear H~II contamination in spectrum.\\
1g: Strong HST [Fe~II] 1.64 $\mu$m candidate\\
1h: HST Nuclear candidates (mostly \citealp{dopita2010}) with no additional supporting evidence.\\
1z: Special cases.\\
2: Optical candidate with imaging or spectroscopic ratio 0.25 < [S~II]:H$\alpha$ < 0.5.  Many of these are likely SNRs, but H~II contamination causes uncertainty.  Some have detectable [Fe~II], which strengthens their case.\\
h: Historical supernova (event was observed).\\

Finally, the catalog has a use flag column, which indicates how each SNR was used or not used in the analysis.  This flag as the following categories:

\begin{itemize}
    \item p: Progenitor mass constrained from measured local young population.
    \item n: No young population detected in the fit (Type Ia candidate).
    \item o: Outside of the HST coverage; no measurement performed.
    \item c: Too close to the center of M83; no measurement performed.
    \item i: Insufficient photometry for a measurement; fitting failed.
\end{itemize}

\begin{figure*}
	\centering
    \includegraphics[scale=0.75]{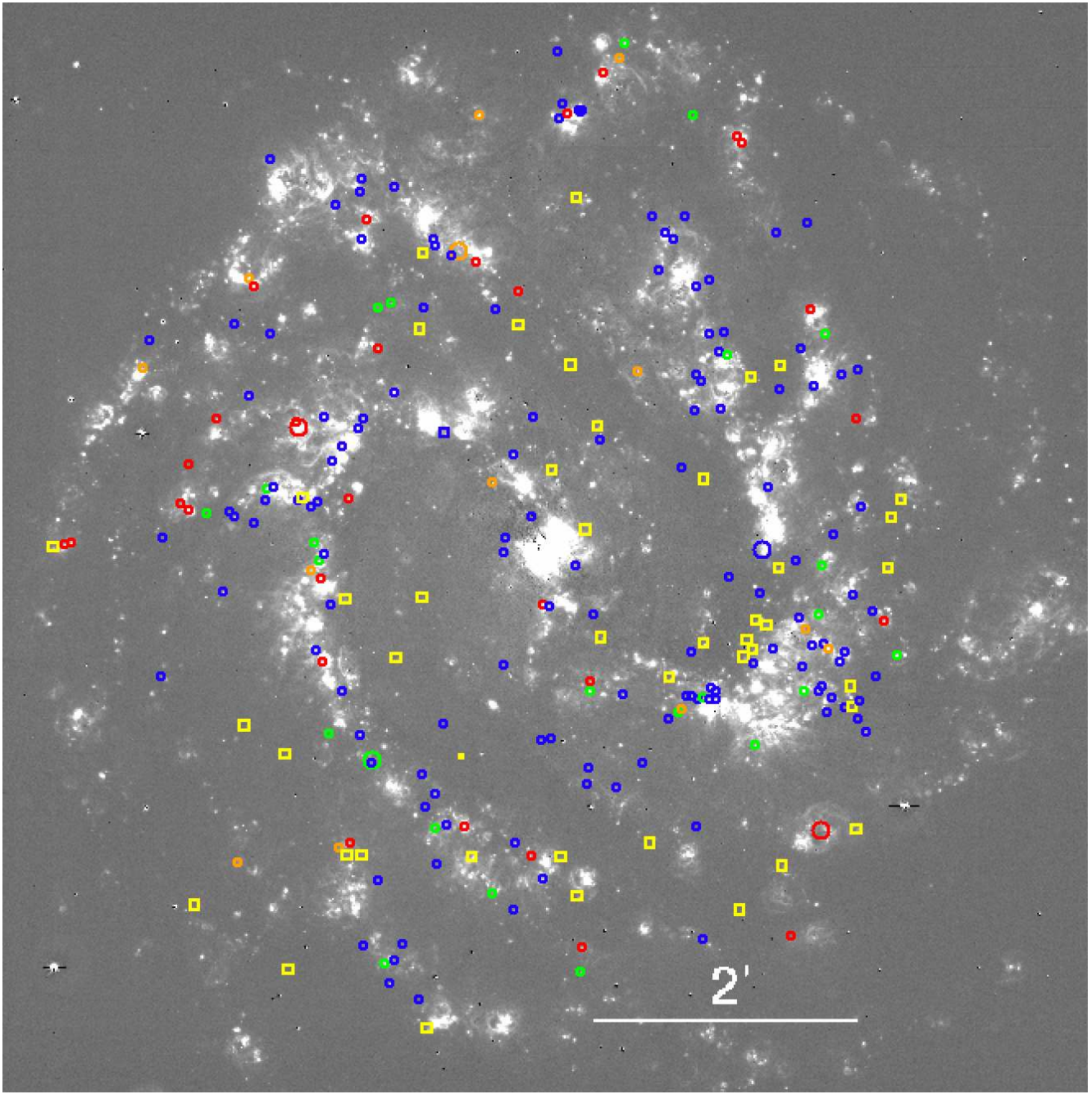}
    \caption{Locations of M83 SNRs color-coded by the progenitor masses inferred from their local stellar populations are overplotted on a continuum-subtracted H$\alpha$ image from the Magellan telescope \citep{blair2012} Colored circles indicate: red = $>$20 M$_{\odot}$; orange = 16-20 M$_{\odot}$; green = 12-16 M$_{\odot}$; blue = $<$12 M$_{\odot}$. SNRs with no local young population (Type Ia candidates) are plotted as open yellow squares.  Slightly larger circles show positions of historical SNe for which we derived progenitor masses. \label{fig:massPositions}}
\end{figure*}

In Figure~\ref{fig:massPositions}, we show the locations of all of the SNRs
for which we were able to obtain sufficient photometry for a
measurement (use flags ``p'' and ``n'').  The circles mark remnants of
use flag ``p'' and are color-coded by the most likely progenitor mass
from our technique.  Yellow squares mare the locations of remnants
with ``n'' use flags.

The SNR catalog has 307 historical SNe, SNRs, and SNR candidates.  There are 271 of these within the 7-field HST/UVIS footprint for which we have resolved stellar photometry.  However, 24 of those locations are too close to the complex galaxy nucleus to provide reliable resolved stellar photometry.  Of the remaining 247 SNRs, 10 more had samples that proved insufficient to allow the fitting to converge. Our final measured sample therefore consisted of 237 SNR locations (see Figure~\ref{fig:massPositions}) and a 100 random control sample positions.  

For our program, we first ran photometry on all of the resolved stars in the HST imaging data.  Then, to quantify the errors and completeness of the photometry as a function of star color, magnitude, and location, we ran over a million artificial star tests.  As in previous population-based progenitor studies, we then fit the photometry data within 50 pc of each SNR with stellar evolution models to constrain the age distribution of the young stars near each SNR.  We now discuss the production and analysis of our resolved stellar photometry.

	\subsection{Photometry}
		
        
    We measured resolved stellar photometry from the F336W, F438W, and F814W imaging with the point spread function (PSF) fitting photometry pipeline used for the Panchromatic Hubble Andromeda Treasury \citep[PHAT;][]{williams2014phat}.  In short, this pipeline uses an astrodrizzle PyRAF routine to flag cosmic-ray affected pixels, and the DOLPHOT \citep{dolphin2000,dolphin2016} PSF-fitting photometry package with the TinyTim-generated PSFs \citep{krist2011} to find and measure point source photometry for all sources in a set of HST images.  It starts by reading all of the {\tt flc} images into memory to search for stars using the full image stack. Then the photometry is performed on the individual {\tt flc} images, where a PSF is fitted to the location of all star locations in every image.  For the multiple exposures in a given filter, the measurements are combined to reduce the photometric uncertainty and increase the signal to noise of the measurements.  
    
    The pipeline has been updated based on some of the detailed tests performed on the PHAT photometry.  In particular, we used the TinyTim PSFs \citep{tinytim} and we used the CTE-corrected {\tt flc} images for photometry instead of post-photometry CTE corrections.  Both of these updates have resulted in more precise photometry with smaller systematic uncertainties. 
    
\begin{figure*}
	\centering
    \includegraphics[scale=0.75]{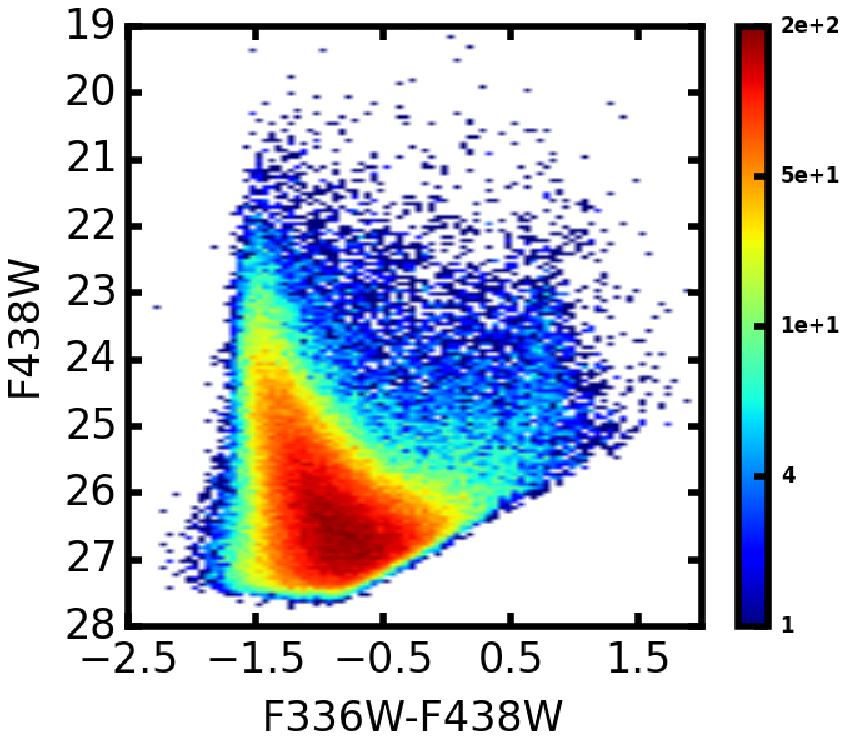}
    \includegraphics[scale=0.75]{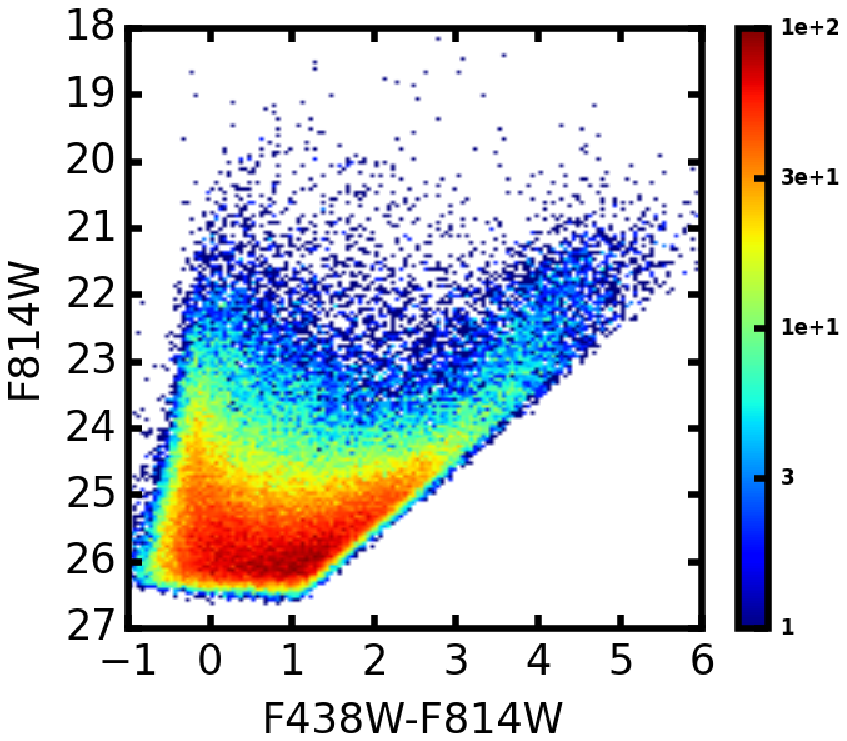}
    \includegraphics[scale=0.75]{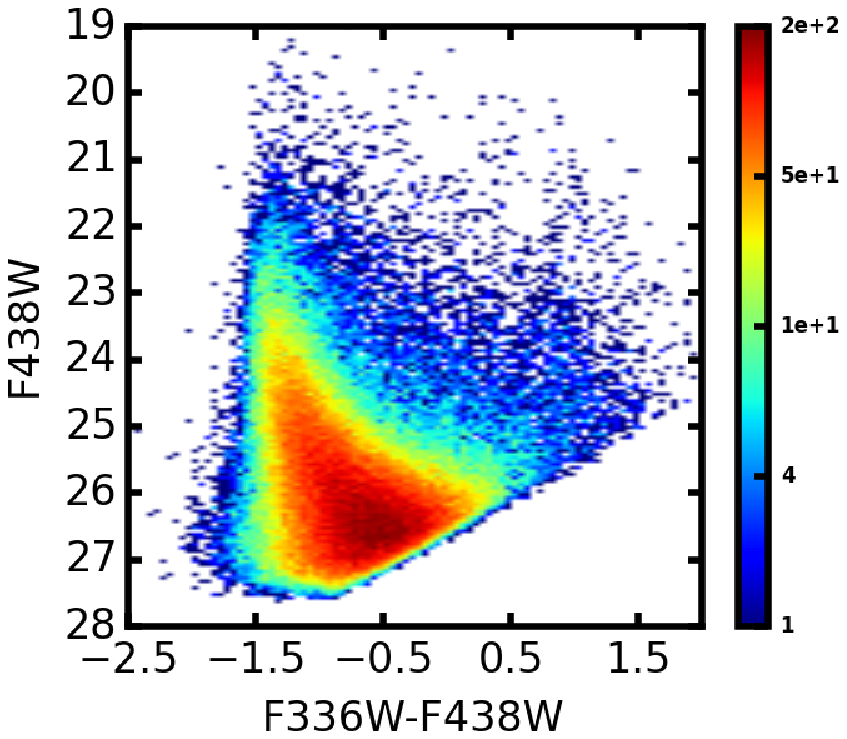}
    \includegraphics[scale=0.75]{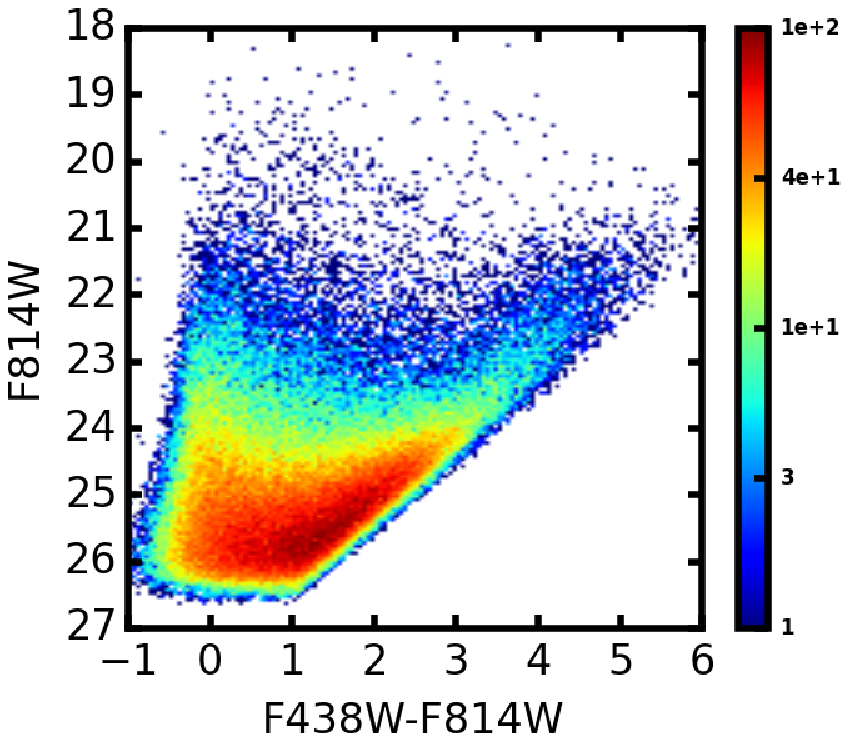}
    \caption{Color-magnitude diagrams for the three UV/Optical bands for two fields of the the seven field mosaic of M83. These are ({\it upper-left}) UV CMD for The POS2 field, ({\it upper-right:}) optical CMD for the POS2 field, ({\it lower-left:}) UV CMD for the F5 field, and ({\it lower-right:}) optical CMD for the F5 field. \label{fig:CMDs}}
\end{figure*}

From the DOLPHOT measurements, we first produced full-field
color-magnitude diagrams (CMDs) of our photometry.  We show examples
of these in Figure~\ref{fig:CMDs}, where it is clear that we have
detected the upper main-sequence of blue stars.  At the distance of
M83, the depth of these CMDs is to m$_{F435W}{\sim}26$;
$M_{F435W}{\sim}-2.5$, which corresponds to a main-sequence turnoff
age of 70 Myr, which is the lifetime of a 6.3 M$_{\odot}$ star in the
Padova models.  Therefore we are sensitive to all populations relevant
for CCSNe progenitor measurements.  

\begin{figure*}
	\centering
    \includegraphics[scale=0.5]{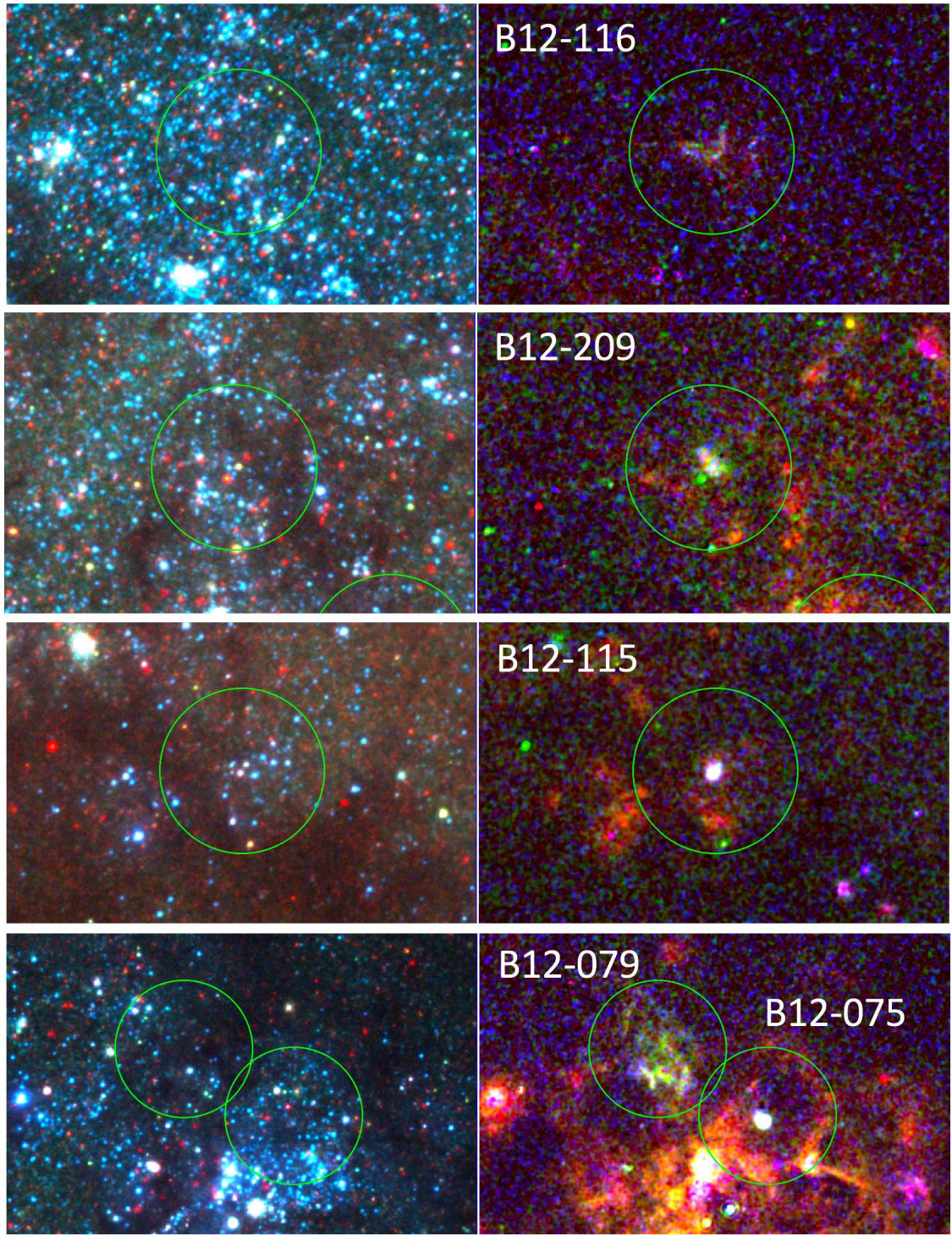}
    \caption{Example fields showing the range of star colors, star density, and impacts from dust absorption encountered in the 100 pc fitting regions around several M83 SNRs.  All data shown are from HST WFC3 imaging and the green circles are 100 pc in diameter at the assumed distance of M83.  Each two-panel pairing shows a three-color image of stellar data on the left (red = F814W, green = F438W, blue = F336W), and a three color continuum-subtracted emission line image on the right (red = H$\alpha$, green = [S II], blue = [O III]).  Identifications for the SNRs are from \citet{blair2012} and are shown on the emission line panels.\label{fig:images}}
\end{figure*}

After this check, we focused our further analysis on the regions
within 50~pc of each SNR location, and on a control set of photometry
taken from random positions of the same physical size.  These control
regions were distributed in proportion to the real SNRs in each of the
7 M83 UVIS fields, and their photometry samples were run through the same analysis routines as those of the SNR regions.  The results from the control samples allow us to
  quantitatively check that the measurements from the SNR locations
  differ from the field.  Examples of the relevant regions around
several SNRs are shown in Figure~\ref{fig:images} to demonstrate how
well-resolved the individual stars are and to show the diversity of
the dust and stellar density properties of the regions.
    
    \subsection{Artificial Star Tests}\label{sec:optimization}

Fitting CMDs with models requires convolving models with the errors,
bias, and completeness of the photometry at each color and magnitude.
These quantities are best determined through artificial star tests
(ASTs), whereby a star of known location, color, and brightness is
added to the images, and then the photometry of that section of the images
is rerun.  This operation is run hundreds of thousands to millions of
times to cover location, color, and brightness space well enough to
properly model the photometric quality of each CMD that we fit.

We ran ASTs covering the range of colors and magnitudes seen in the
observed stellar catalogs.  The differences between the input star
properties and the output measurements allow us to calculate the bias,
uncertainty, and completeness of the observed stellar catalog as a
function of star color, and magnitude, and environment.  These quality
metrics are then used by the fitting software to convolve model CMDs
so that they have the same photometric quality as the observed CMDs.
Then we can determine the model CMD that best matches the observed
CMD.
    
ASTs require significant computational resources as they require
running the photometric measurements many times.  Therefore, to
increase the efficiency of ours, we assume regions of similar stellar
density and exposure time will have resolved star samples of the same
photometric quality.  This assumption allows us to apply one set of
ASTs to several SNRs.  A similar technique was used by the PHAT team
to drastically decrease the number of ASTs necessary to measure SFHs
as a function of position over the large area covered by the PHAT
project \citep{williams2017}.

To optimize the efficiency of our ASTs, we first measured the relative stellar density as a function of
  position in the images.  To avoid biases due to completeness
  differences we limited the magnitude range over which we counted
  stars to where the completeness was high ($>$80\%) over the entire survey (F438W<26.2).  At each image location, we counted stars
  above this brightness limit within 50~pc and
  divided by the area to determine the stellar density.

We then tested the assumption that ASTs taken from regions of
  similar stellar density are interchangeable.  For this test, we ran
  ASTs at the location of an observed stellar sample, as well as in
  another location in the image with similar stellar density.  We then
  fit the observed CMD of one of the locations twice: once with each
  set of AST results.  We found that ASTs from locations with stellar
  densities within a factor of two of one another resulted in
  equivalent age measurements.  Thus, we performed 42 sets of ASTs
  obtained from regions spanning the full range of stellar densities
  in our sample (0.4 to 20 arcsec$^{-2}$), and stay well within this
  factor of two variation. We then applied this library of ASTs during
  model fitting to our photometry.  Namely, when fitting each SNR and
  control sample, we convolved the models with the photometric errors,
  bias, and completeness by applying the ASTs measured from a region
  of appropriate stellar density.

    
    \subsection{Deriving Star Formation Histories}
    
We employ the CMD fitting package MATCH
\citep{dolphin2002,dolphin2013}, which fits the CMDs
from resolved stellar photometry to determine the best star formation
history (SFH) of each stellar sample near a SNR and control
  region. The SFH is the distribution of stellar ages and
metallicities present in the sample, as well as the uncertainties on
that distribution.  This package has been used by many investigators
to measure the star formation rate as a function of lookback time for
a number nearby galaxies \citep[e.g.,][and many
  others]{williams2009,mcquinn2010,weisz2011,skillman2017}, as well as
to measure star cluster ages \citep[e.g.,][]{senchyna2015}, and the
ages and masses of supernova progenitors
\citep[e.g.,][J14]{2011ApJ...742L...4M,jennings2012}.

We applied time bins starting at 6.6 to 8.0 log years in steps of 0.05 dex and then from 8.0 to 10.1 log years in steps of 0.1 dex using Padova isochrones \citep{girardi2002,2008A&A...482..883M, girardi2010}.  Furthermore, we constrain the present day metallicity from $-0.6{\leq}[Fe/H]{\leq}0.1$, as appropriate for M83 \citep{bresolin2009}.   We also applied reddening ($A_V$) and differential reddening ($dA_V$) to our model fits (see Section 2.5).  In addition, the MATCH package allows the inclusion of a contamination CMD to look for populations that are present in the sampled region at higher density than the surrounding field.  We used the entire WFC3 162$"{\times}162"$ UVIS M83 field surrounding the SNR, scaled to the area of the extraction region area, for this contamination CMD.

\subsection{Determining foreground and differential extinction}

M83 is characterized by an abundance of dust embedded in the disk. For extinction, it is valuable to consider multiple colors. Therefore, we consider both the F336W-F438W and F438W-F814W CMDs when finding the extinction properties of each location.  These extinction properties are both the overall extinction affecting the entire location, $A_V$ (with the minimum bounded by the foreground Galactic extinction of 0.2; \citealp{2011ApJ...737..103S}), and extinction spread due to the distribution of stars along the line of sight ($dA_V$).  Each of these parameters affect the CMD differently, and are therefore they are applied to the models independently.  First, $A_V$ is applied to all of the stars in each model CMD.  Then $dA_V$ in applied to the model CMD to spread the stars along the reddening line as they would be if they were randomly distributed within a uniform dust layer. In essence, $A_V$ affects the location of features in the CMD (e.g., the main sequence), while $dA_V$ affects the sharpness of the features in the CMD.  For example, stars embedded within dusty regions of a galaxy (high $dA_V$) will produce CMDs with features smeared along the reddening line, even if they are behind little foreground $A_V$, while stars not embedded in a dusty region (low $dA_V$) will produce CMDs with sharp features, even if all of the stars behind a large foreground $A_V$.

We determine the values for $A_V$ and $dA_V$ by finding the best maximum-likelihood fit value reported by MATCH \citep{dolphin2012} for a grid of possibilities when fitting (see Figures  \ref{fig:m215_measurement} and \ref{fig:m051_measurement}).  We then fix the  $A_V$ and $dA_V$ values to those determined from the multiple colors when fitting the deepest single CMD for the final age distribution and uncertainties out of these, as described below.
            
\subsection{Uncertainties\label{sec:uncertainties}}

Our uncertainties are mainly due to photometric error, incompleteness, and the number of stars in our sample available to constrain the age distribution of the young component of the population.  All of these vary significantly for the SNR sample.  For example, the number of stars ranges from 10s to 100s of stars per SNR.  Thus we measure the uncertainties separately for each SNR sample. 

We characterize these uncertainties using a hybrid Monte-Carlo (MC) algorithm within MATCH using the task \texttt{hybridMC} \citep{dolphin2013}.  This task accepts or rejects thousands of potential SFHs surrounding the best fitting one, based on likelihood.  It then applies the distribution of accepted SFHs to determine the 68\% uncertainty regions surrounding the best fit.  The \texttt{hybridMC} uncertainty determination applies statistics that are valid only for independent CMDs.  Applying these statistics to 2 CMDs that share a filter would produce incorrect uncertainties because the CMDs are correlated. Therefore, we must choose one of our CMDs when fitting for the final SFHs and uncertainties.   The M83 data contain imaging in F336W, F438W, and F814W. 

To determine which CMD fit to use for our final SFH and uncertainties, we fit both the F336W vs. F336W-F438W and the F438W vs. F438W-F814W CMD independently, fixing $A_V$ and $dA_V$ measured values (see Section 2.5).  We then compared the best-fit SFH to the best-fit SFH from the simultaneous fit of both CMDs.  We found overall consistency between the techniques, but the single CMD fits had reliable uncertainties.  The best constraints were almost always from the fits to the  F336W vs. F336W-F438W CMDs.  This result was expected because the F814W photometry is the shallowest and least sensitive to the young populations. However, there were a handful of cases, mostly those with the highest reddening, where the F336W-F438W CMD alone did not produce a result consistent with the 2 CMD fit from the $A_V$, $dA_V$ runs (185, 278, 059, 197, 090, 117, 150, 227, 246, SN1950B).  In these cases we found that the fit to the F438W vs. F438W-F814W CMD produced a result consistent with the 2 CMD fit.  We report as our final measurement of the SFH and uncertainties, the results from the F336W vs. F336W-F438W CMD, except for these exceptions.

 \subsection{Progenitor Mass\label{sec:progenitorMass}}
 
Our method for constraining the progenitor mass is similar to that of previous work (e.g.,\citealp{jennings2012}; J14; \citealp{williams2018}). The method begins by finding the most likely age of the progenitor star.  In short, we measure the SFH from broadband photometry of the resolved stars within 50 pc of the location of each SNR.   This region size assumes that most young stars are coeval within our desired extracted region, as suggested by the work of \citep{2006MNRAS.369L...9B}.  Indeed, the validity of this assumption has been strengthened by the works of \citet{2009ApJ...703..300G} and \citet{2011ApJ...742L...4M}.
            
We restrict the ages in the SFH that we apply to our progenitor analysis.
The ages of SNR progenitors measured with deeper photometry in more nearby galaxies have established that the maximum progenitor age for CCSNe is $<$50 Myr (young enough for a $\sim$7~M$_{\odot}$ star to still exist).  While older progenitors are allowed according to binary evolution models \citep[e.g.,[]{xiao2019}, a maximum of 50~Myrs is empirically measured for CCSNe progenitors by J14 and confirmed by   \citet{diaz2018} in M31 and M33, even when they include older ages in their analysis.  Furthermore \citet{xiao2019} find a similar empirical age constraint using an independent emission line fitting technique that uses the latest binary evolution models to infer ages of HII regions that have hosted CCSNe.

We take advantage of these empirical results to limit the part of the SFH that we apply to determine the progenitor mass.  We assume only the part of the SFH $<$50~Myr is relevant for determining progenitor ages.  When we infer the most likely progenitor age of each SNR, we look for the most prominent peak in the age distribution, only considering ages $<$50~Myr.  This assumption means that we are not able to use these data to probe the high-age (low-mass) end of the progenitor mass function, but it allows us to avoid including the higher age bins, which have large uncertainties at the photometric depth of the observations used here, into our age determinations.

We derive the mass fraction for each time bin over the past 50~Myrs, taking advantage of the empirical constraints so that we do not need to include older ages that have main sequence turnoffs near our completeness limit which lead to large SFH uncertainties.  Furthermore, these older bins are more likely to contain stars unrelated to the SNR progenitor because stars of older ages have had more time to migrate away from their siblings.  Thus, limiting the ages to those known to be relevant to CCSNe progenitors helps to minimize contamination from unrelated populations.

\begin{figure*}
	\centering
    \includegraphics[scale=0.35]{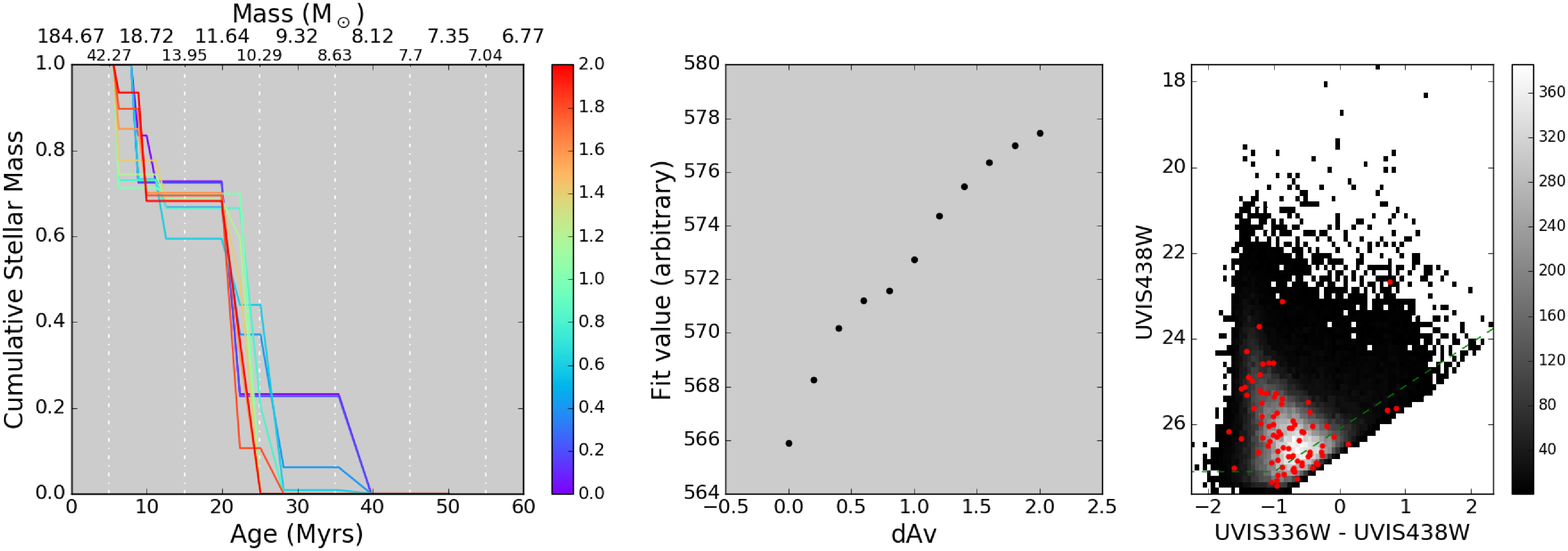}
    \includegraphics[scale=0.2,trim={4cm 0 1cm 0},clip]{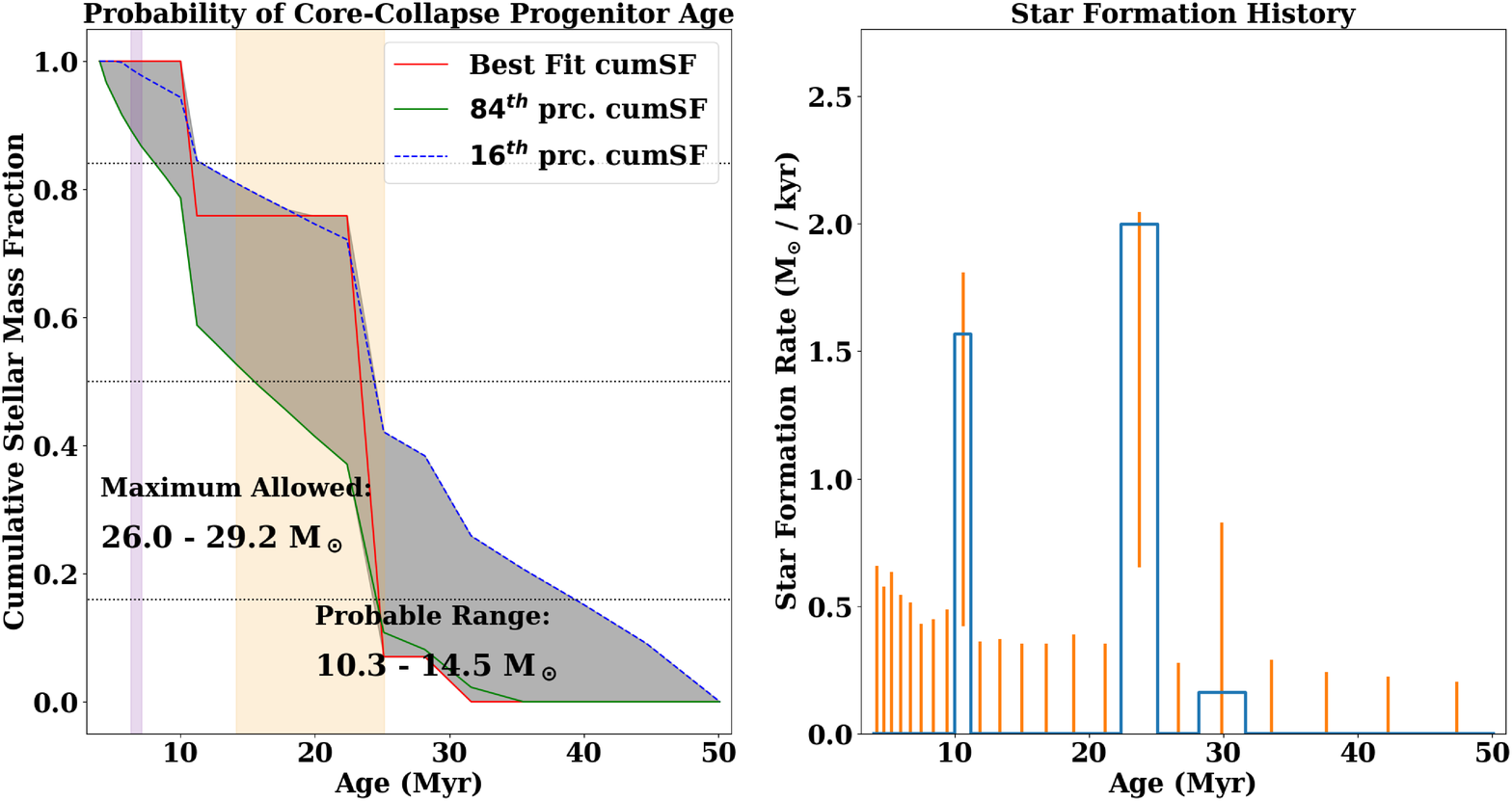}
    \includegraphics[scale=0.18,trim={2cm -4cm 0 0}]{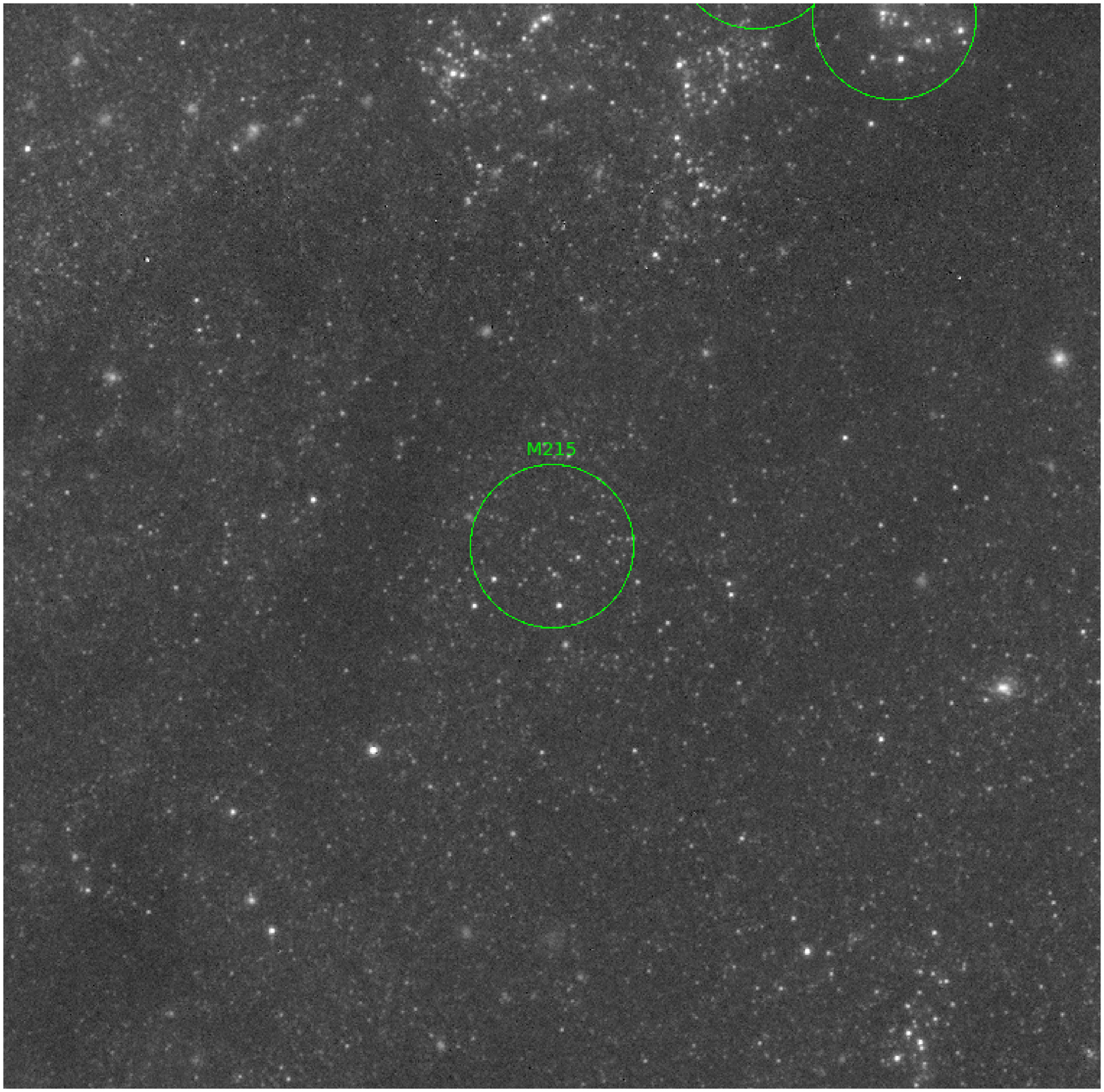}
    \caption{Technique for constraining SNR progenitor mass (SNR~295). {\it Bottom right:} The extraction region for the resolved stellar photometry sample is shown as a green circle on a 20$''\times$20$''$ F438W image of the region of interest for this SNR. {\it Top right:} The observed CMD from the extraction photometry.  The photometry of the stars in the surrounding field are shown in grayscale, where the lighter colors mark regions of the CMD that have more field stars, as indicated by the color bar.   The photometry of the stars within 50 pc of the SNR location are overplotted in red points.  The magnitude limits of the data included in the CMD fitting are shown by the dashed lines.  These limits are are based on the completeness for regions with the stellar density of this locatio,n as determined by the ASTs (see \S~\ref{sec:optimization}). {\it Top left:} Multiple SFH measurements, assuming different amounts of differential reddening (dAv) at the SNR location. Lines show the cumulative fraction of stellar mass in each age interval.  Various colors denote the amount of dAv, as indicated by the color bar. {\it Top center:}  Maximum likelihood fit values as a function of assumed dAv.  Lower values indicate a better fit to the data.  In this example, we adopt a dAv value of zero for the final fitting, as this value produces the best fit. {\it Bottom left:}  The cumulative fraction of stellar mass in each age interval for the final SFH fit, along with uncertainties from the {\tt HybridMC} error analysis. The probable mass range listed is the range of ages consistent (1$\sigma$ uncertainty) with the population median age, marked with the tan shading. The maximum allowed progenitor mass is the most massive star allowed in this location by the uncertainties in the youngest age bins. {\it Bottom center:} The differential SFH, showing the star formation rates and uncertainties that correspond to the cumulative fraction plot in {\it bottom left}.  Note that bins with large uncertainties have a large covariance with neighboring bins, which makes the cumulative distribution easier to interpret.  {\it We include figures of this format for all 199 SNRs with progenitor constraints in the supplemental materials.
}  
    \label{fig:m215_measurement}}
\end{figure*}

\begin{figure*}
	\centering
    \includegraphics[scale=0.35]{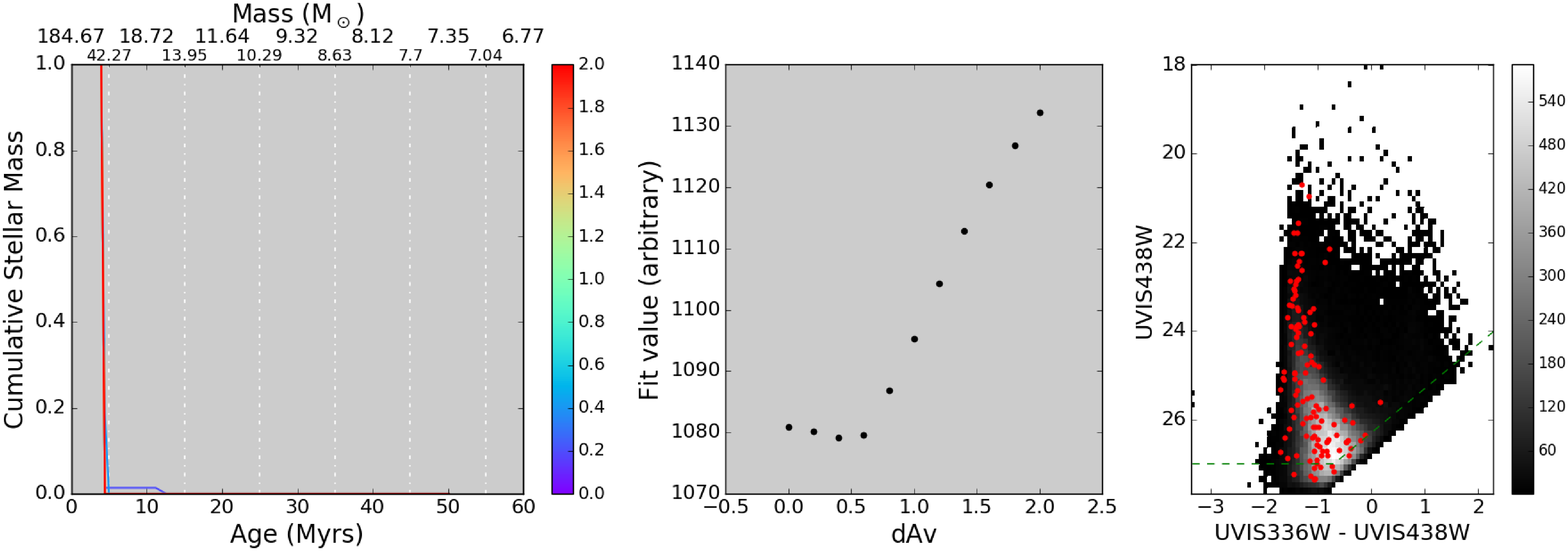}
    \includegraphics[scale=0.2,trim={4cm 0 1cm 0},clip]{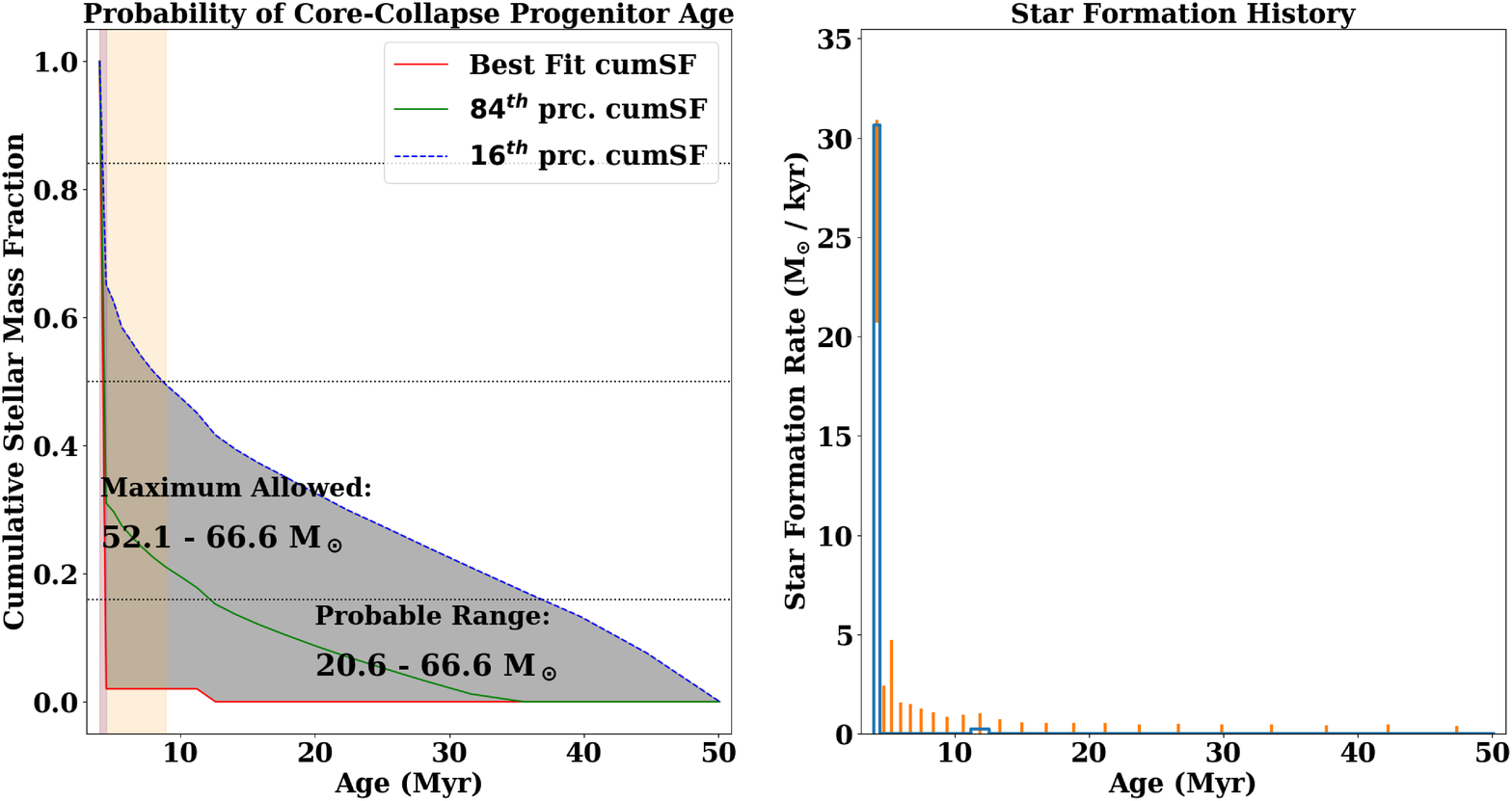}
    \includegraphics[scale=0.18,trim={2cm -4cm 0 0}]{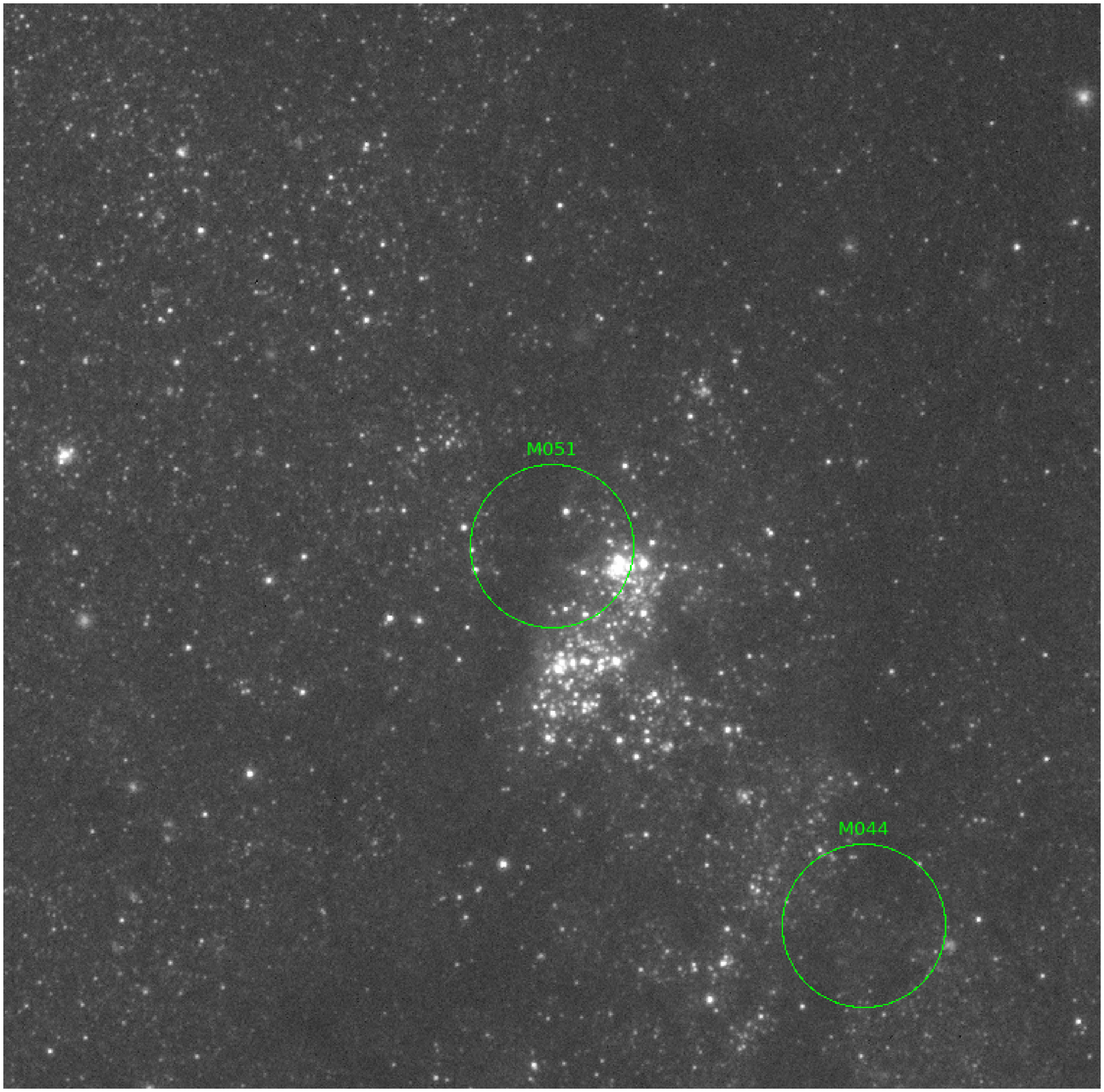}
    \caption{Same as Figure~\ref{fig:m215_measurement}, but for 057.  We provide this second example to show an extreme case, as 057 has the youngest age constraint of the entire sample. \label{fig:m051_measurement}}
\end{figure*}

Figures~\ref{fig:m215_measurement} and \ref{fig:m051_measurement} summarize the progenitor mass measurement technique for two SNR locations (SNR~295 and SNR~057, respectively).   We include figures of this format for all 199 SNRs with progenitor constraints in the supplemental materials.   The effect of the assumed differential extinction is shown in the upper-left panel, where the resulting age distribution from each assumed differential extinction amount is plotted with a different color.  In Figure~\ref{fig:m215_measurement} the general pattern in ages is relatively insensitive to the assumed $dA_V$. The lower-left panels of the figure show the star formation history (SFH) along with uncertainties for the best $dA_V$ result.  Peaks in the rate and stellar mass fraction correspond to the most likely progenitor age. 

The full SFHs show how well the median age represents the full age distribution of the population.  An age distribution may correspond to several epochs of SF or may be dominated by one epoch.  In the case of the former, the median age is less likely to represent the correct age of the progenitor.  The full probability distribution for each event is a more precise characterization of the complex uncertainties.
           
Based on these measurements, we use the age where 50\% of the stars (as determined from the bottom-center panel) have formed as the most likely progenitor age.  We then take our uncertainties to include all ages that may contain the median age.  This is shown with the tan vertical shaded region in the bottom left panel of the mass measurement figures (e.g. Figures~\ref{fig:m215_measurement} and \ref{fig:m051_measurement}).  These limits define our 1$\sigma$ progenitor age range for the progenitor mass.  From this age range, we infer the zero age main sequence (ZAMS) mass range using the \citet{girardi2010} isochrone library as a lookup table for the most massive star present at any given age.   We set a floor for the uncertainty to be at least 0.1~M$_{\odot}$.  In the cases where we find there to be zero SF in the last 50 Myrs, we assume these to be the remnants of Type Ia supernovae or runaway massive stars, and we provide no measurement of $M_{ZAMS}$.

\section{Results}
     
 After completing our measurements of the recent SFHs surrounding all of the SNRs and converting these to probability distributions, we generated Table \ref{tab:report} describing all SNRs including the masses corresponding to the median ages surrounding each.  In column 1 we give the identifier of the SNR. Columns 2 to 4 specify the 50\% completeness magnitudes for each SNR for the three filters in our analysis.  Note these cutoffs are derived from the 50\% completeness of the respective fake star sets; hence, SNRs utilizing the same fake stars will have the same magnitude cutoffs (see \S~\ref{sec:optimization}).  Column 5 gives the number of stars extracted from each region in the F438W data.  Columns 6 and 7 give the best differential reddening and reddening applied during the fitting process, respectively. Lastly, column 8 gives the inferred $M_{ZAMS}$ from the median age with bounds.  When an SNR is missing values in column 8 this means either we were not able to measure the SFH or we found there to be no SF in the past 50~Myrs (see Table~\ref{tab:catalog} use flags n, o, c, and i).  Within this table we only include SNRs where we were able to put a constraint on the progenitor mass (use flag "p"). 
      
In Table \ref{tab:probDistro} we show the calculated age probability distribution for an SNR as the fraction of stellar mass present in each age bin.  The supplemental materials contain similar information for all of the SNRs we measured.  In columns 1 and 2 we define the edges of the age bin (T1 and T2) with T1 being the more recent look back time. Column 3 gives the best-fit star formation rate, SFR(best), for that age bin while columns 4 and 5 are the negative and positive uncertainties on that rate.  Column 6, PDF(best), gives the fraction of the $<$50 Myr stellar mass in that age bin according to the best-fit SFH, and  columns 7 and 8 provide the uncertainties on that fraction, given the rate errors in Columns 4 and 5.  Column 8, CDF(best), gives the running cumulative fraction of stellar mass as a function of time for the best-fit SFH. Columns 9, 10, and 11 provide the 16th, 50th, and 84th percentiles of the cumulative mass fraction for a set of 10$^6$ realizations of the SFH with the uncertainties in Columns 4 and 5. Columns 12 and 13 show the masses of stars with lifetimes that match the edges of the age bin (M1 and M2).  Table \ref{tab:probDistro} corresponds to Figure~\ref{fig:m215_measurement} which has multiple epochs of SF according to the bottom-middle and bottom-right panels.  We see that the progenitor is more likely to belong to the peak at 25 Myr.  
      
\section{Discussion}	

We now investigate the total distribution of masses in our sample as well as looking for spatial correlations between progenitor mass and galactic structure.

\subsection{Spatial Distribution As Function of Mass}
       	
The derived progenitor masses are plotted on an image of M83 in Figure \ref{fig:massPositions} to show the spatial distribution as a function of derived progenitor mass.  The symbol colors denote derived mass with masses above 20~$M_\odot$ shown as red, and lower mass ranges progressing from orange to green to blue.  Yellow squares designate those SNRs for which no young stellar population was identified, and these are the candidates for SNIa progenitors. The higher mass progenitors generally follow the spiral arms, but we note that the most massive progenitor masses appear to be most closely associated with arms and/or star forming regions.  This result is consistent with findings in clustering studies that show trends where stars of higher masses are more clustered \citep[e.g.,][]{kang2009,bianchi2012,kim2012}.  Lower mass progenitors, while also associated with the arms and star forming regions, seem to show a broader spatial distribution.  We also see that the Type Ia candidates tend to be even more generally distributed, including the interarm regions.

\subsection{Progenitor Mass Distribution}

\begin{figure*}
	\centering
    \includegraphics[scale=0.75]{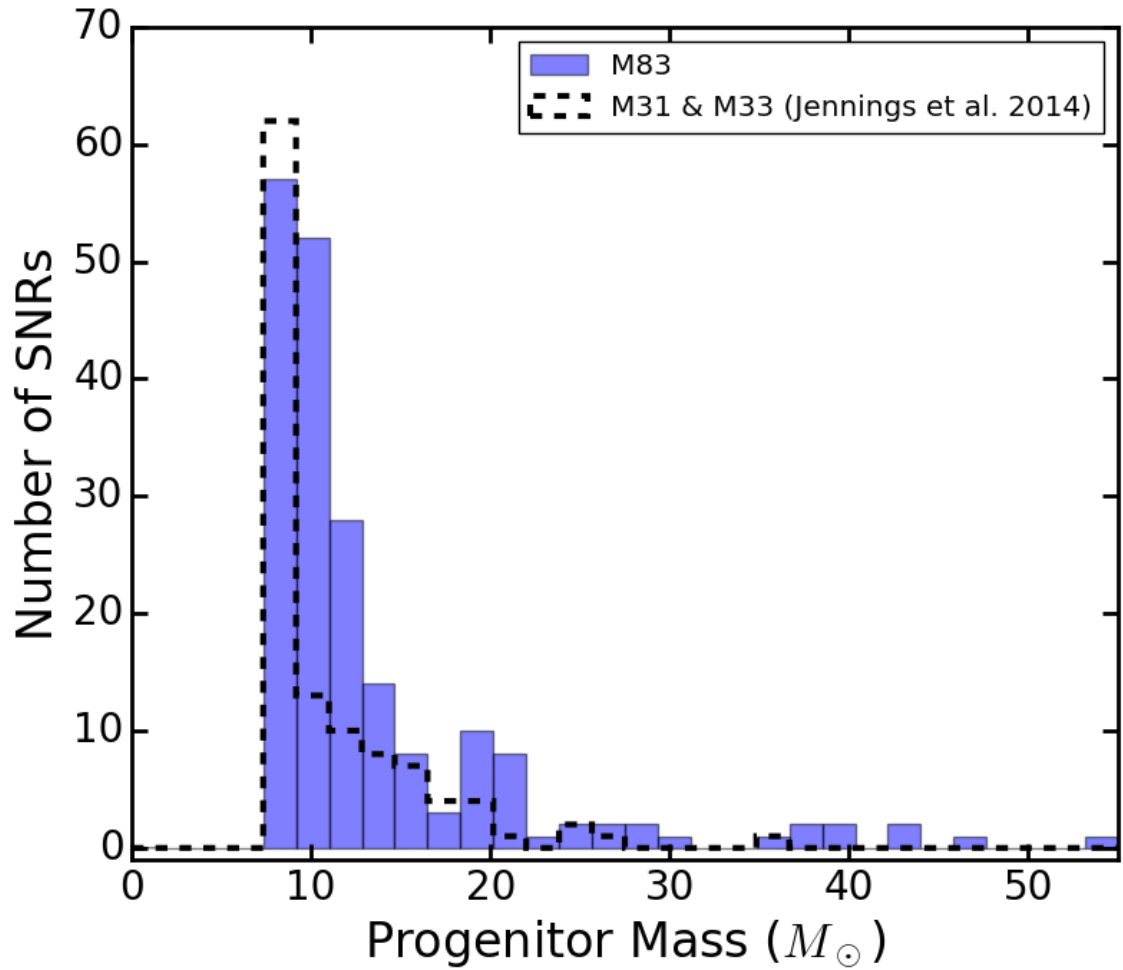}
    \caption{The M83 distribution of progenitor masses is shown by the blue histogram, with the M31+M33 distribution (dashed line) for comparison.  The M83 sample has many more SNRs with measurements at masses $>$10 M$_{\odot}$ but is not as dominated by the lowest mass bin.  M83 also has a significant tail to higher masses, with a number of bona fide progenitor masses in excess of 30 M$_{\odot}$. The cutoff at lower masses was actually measured for the M31+M33 sample \citep{diaz2018}, but was assumed for this work on M83 (see Section~2.7 for details). \label{fig:distro}}
\end{figure*}

\begin{figure*}
    \includegraphics[scale=0.3]{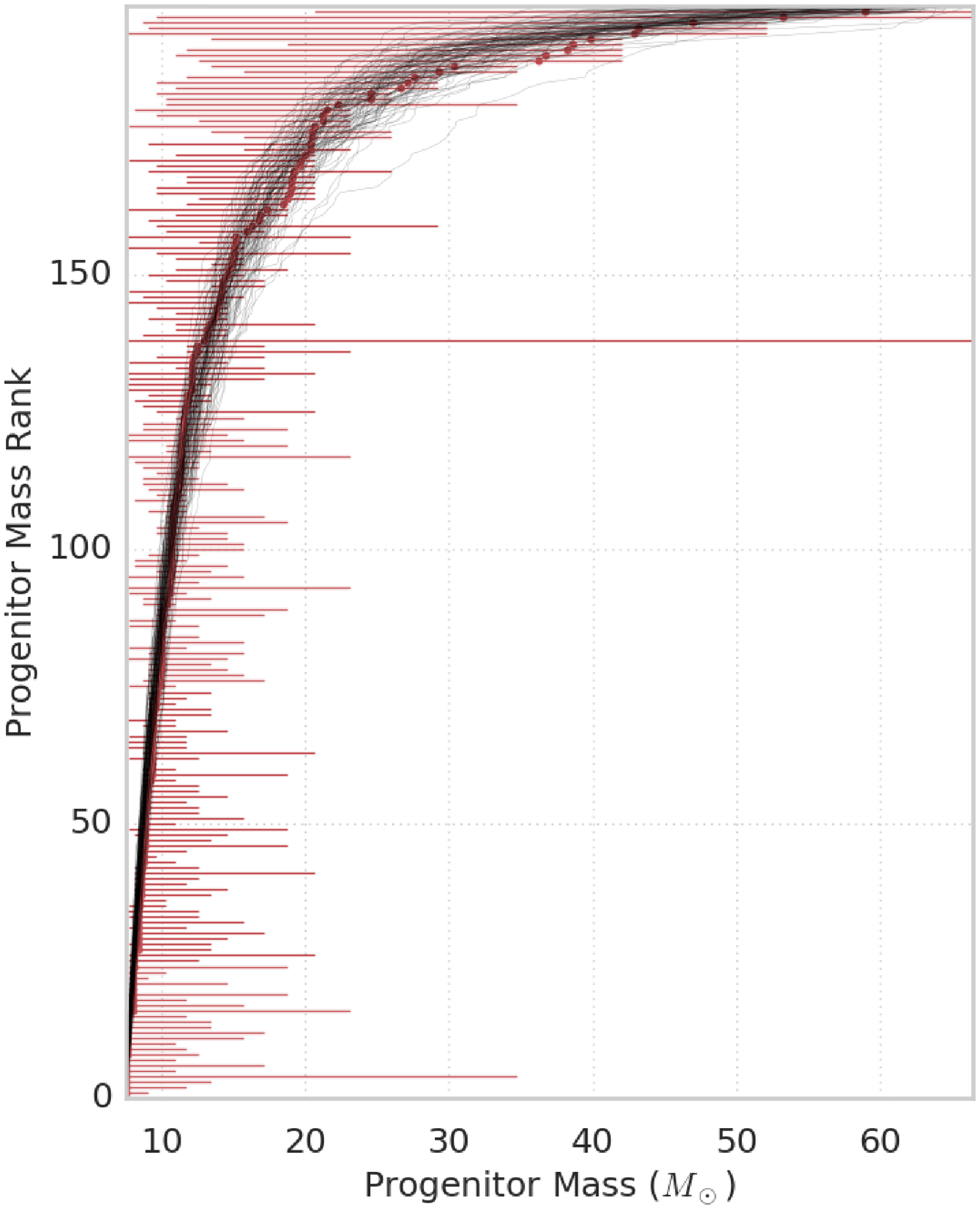}
    \includegraphics[scale=0.3]{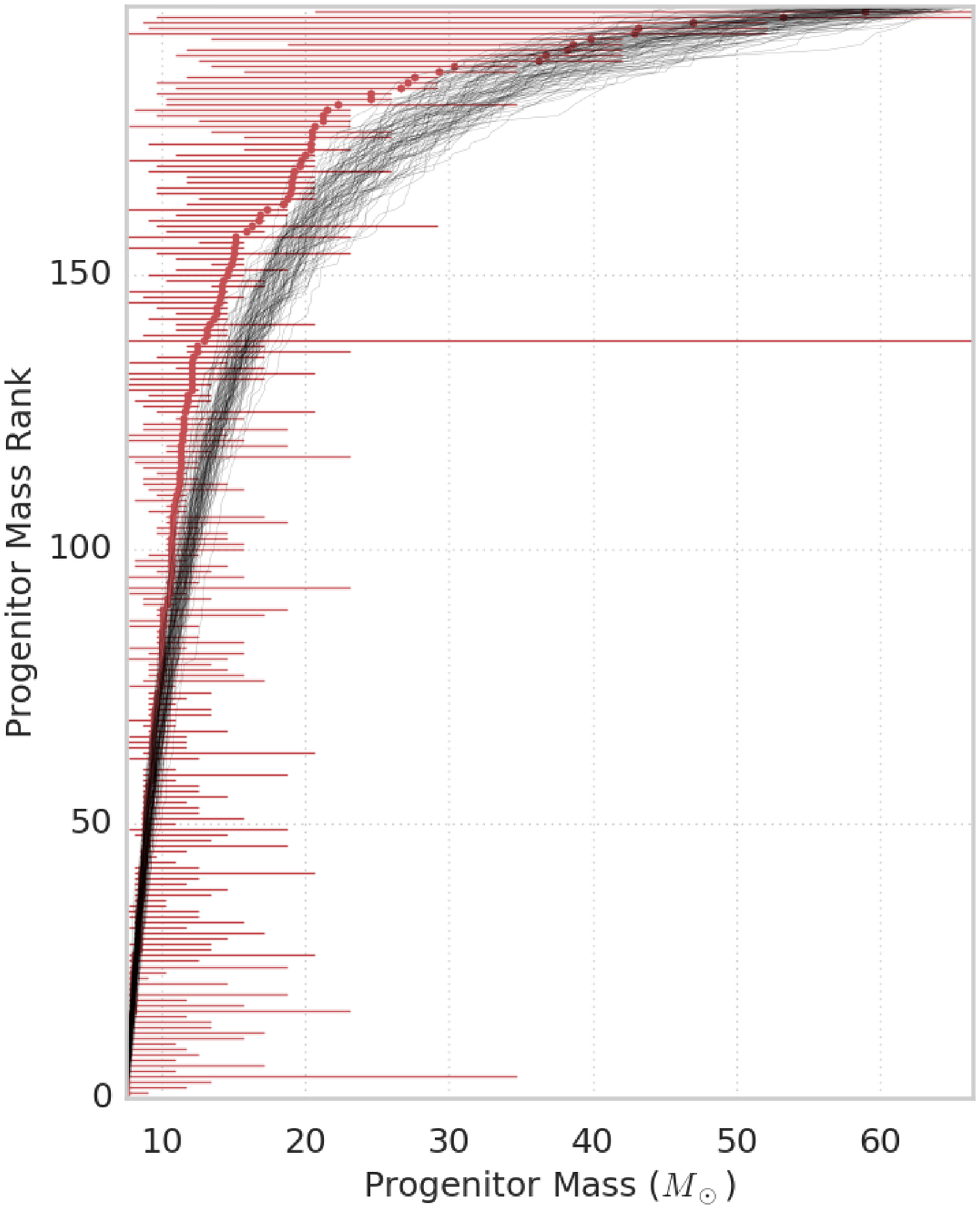}
    \includegraphics[scale=0.3]{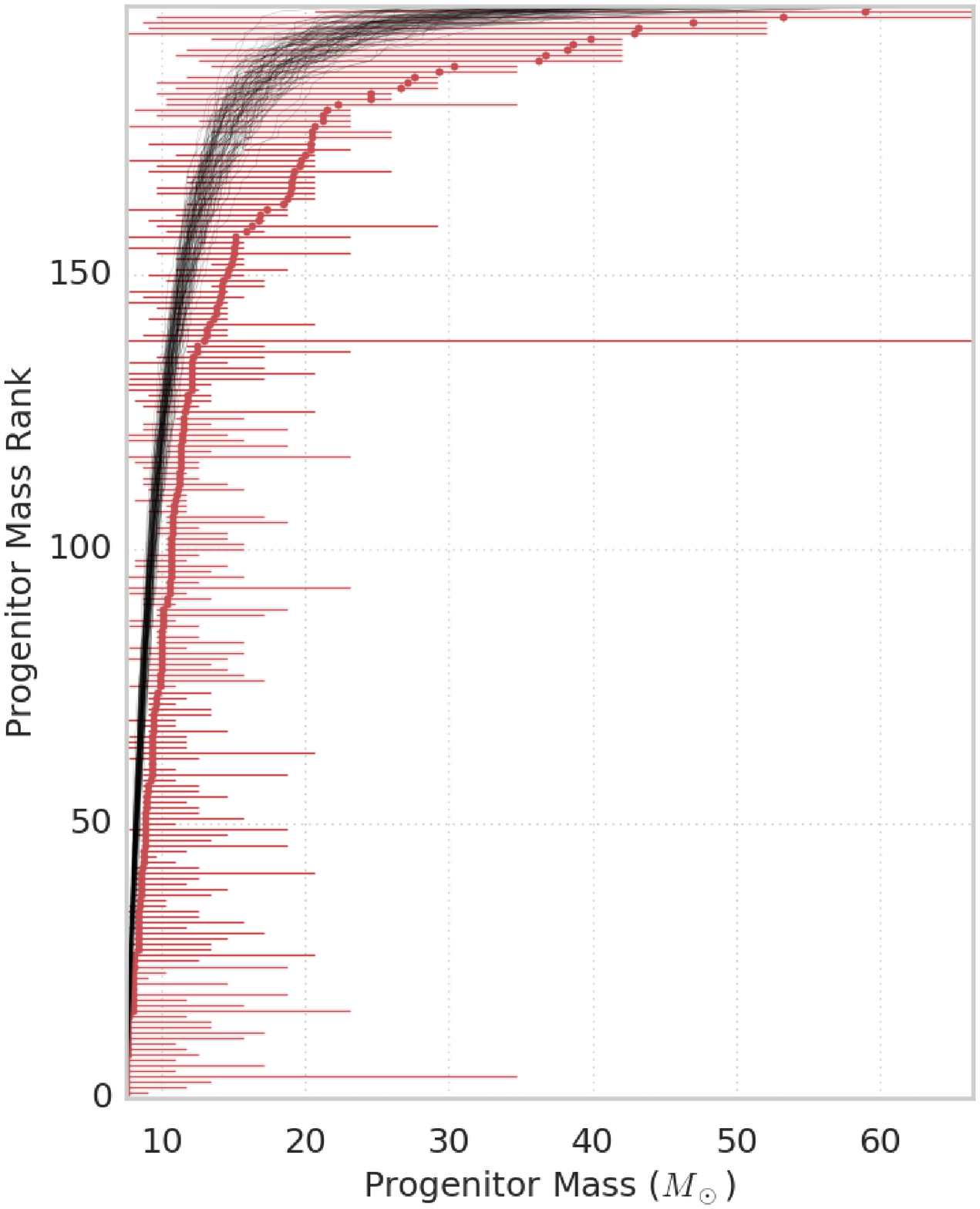}
  	\caption{The ranked distribution of progenitor masses for the 199 measurements we obtained from from the M83 sample are plotted with red horizontal error bars.  Overplotted with gray lines in each panel are 50 draws from a power-law distribution.  While all 3 panels show the same progenitor masses in red, the power-law plotted in gray is different for each panel as follows:  {\it Left:} The best-fitting power law index of -2.9$^{+0.2}_{-0.7}$; {\it Middle} A Salpeter index (-2.35);  {\it Right:} The best-fit index from the combined M31/M33 SNR progenitor sample measurements from (-4.4; J14). \label{fig:imf}}
\end{figure*}

Our sample of  199 SNe and SNRs with inferred progenitor masses is the largest ever produced for any single galaxy.  Thus the distribution has the potential to yield the tightest constraints on the shape of the progenitor mass distribution to date.  We plot the distribution of all the progenitor masses in Figures~\ref{fig:distro} and \ref{fig:imf}.  Figure~\ref{fig:distro} shows a histogram of the progenitor mass distribution for M83 along with the combined M31+M33 distribution for comparison.  While the number of progenitors in the lowest mass bin is comparable in Figure~\ref{fig:distro}, the fraction of the total in that lowest bin is less, with significantly more progenitors in the 10 -- 15 M$_{\odot}$ range and near 20 M$_{\odot}$. The M83 distribution has a significantly higher proportion of higher-mass progenitors as well.

In Figure \ref{fig:imf}, the data points mark the masses at the median ages as described in \S~\ref{sec:uncertainties}, along with their uncertainties.  They are plotted in order of mass to show the cumulative mass distribution.  Lines mark fits from various power-law mass distributions.  The left panel is the best-fit to the data, with an index of $-3.0^{+0.2}_{-0.7}$ ($\chi^2/dof{<}1.1$).  The middle panel has lines drawn from a Kroupa IMF (index of 2.3), and the right panel has lines drawn from the index 4.4 that fits the M31+M33 SNR distribution from J14. Note that this plot excludes the assumed Type Ia supernovae, or objects where no young stellar population was detected.


In addition, we can study only the spectroscopically-confirmed SNR sample (quality flag 1a in the Table~\ref{tab:catalog}), along with the observed SNe, which comprises a subsample of 40.  The mass distribution of this subsample has a best-fitting power-law index of 2.8, making it slightly closer to a standard IMF.  Moreover, a more recent re-analysis of the M31+M33 progenitor distribution that uses Bayesian inference to better account for the uncertainties, finds a power-law index for that distribution of $-3.0^{+0.3}_{-0.5}$ \citep{diaz2018}, which is consistent with this M83 result.

We also compare the SNR distribution to the control sample.  The first difference is that the random positions have double the fraction of locations with no young population.   In addition, the distribution of inferred masses for the random positions is significantly steeper, with a best-fit power-law index of $-$3.7.  The SNR locations appear to be providing a distribution different from the general field.  The fact that the random field is steeper and that the spectroscopically-confirmed sample is less steep than the full sample suggests that contaminants in the full sample may be pushing the best fitting power law index higher.

We apply a Kolmogorov-Smirnov test against the mass distribution of M31+M33 J14, and obtain a P-value of 0.02, suggesting only a 2\% chance of them having the same parent distribution.  Indeed, visually, there is a noticeable difference particularly around $\sim10$~$M_\odot$ and with more massive progenitors.  Thus, the M83 SNR sample really does appear to have more high mass progenitors than those in M31 and M33.
        
While Figure \ref{fig:distro} shows that the M83 distribution is more top-heavy than the M31 and M33 samples, it is still not consistent with a standard IMF \citep{kroupa2001}; there are too few high mass progenitors for the number at the lower mass end.  This result may be due to biases in our technique against the youngest ages dominating the local stellar mass. For example, SNe that occur within very young associations, superbubbles, or highly photo-ionized regions may not leave visible SNRs, thus biasing our survey based on optical SNRs.  However, there is also a possibility that some fraction of the most massive stars do not explode as SNe.  This possibility seems to be corroborated by the discovery of a vanishing red supergiant in NGC~6946 with no observed SN \citep{adams2017,vanisher}.

It does appear to be the case that at least some stars $>$30~M$_{\odot}$ explode as SNe.  One such high mass progenitor is that of the SNR 057 (see Figure~\ref{fig:m051_measurement}). This case is the highest progenitor mass constraint to date from population fitting techniques.  The surrounding population in this case is limited almost entirely to the youngest ages for which we have models, making the progenitor highly likely to have been $>$20~M$_{\odot}$.  This single measurement provides some of our strongest observational evidence to date that such massive stars produce SNe, suggesting that any shortage of progenitor masses $>$20~M$_{\odot}$ is more likely to be attributable to a bias against young ages or some massive stars failing to explode, rather than to a hard high end cutoff in the progenitor mass distribution.  Furthermore, there has been some evidence from stellar population analysis that such high mass progenitors may be more likely to produce stripped envelope SNe \citep{maund2018}, making SNR 057 a candidate remnant from a stripped envelope event.

\subsection{Historical Supernovae}

The M83 sample includes six historical supernovae, where the actual explosion was observed.  These include SN1923A, SN1945B, SN1950B, SN1957D, SN1968L, and SN1983N.  For these, the types are known to be II, unknown, unknown, unknown, II, and Ib, respectively.  Thus, the three known types were all core-collapse SNe, which is not surprising given the high star formation rate of M83.  Of these, SN1968L occurred deep in the complex nuclear starburst region where its parent population is not resolved even at HST resolution, but the high star formation rate in this region favors a core-collapse event.  For SN1923A, we find a likely massive progenitor (53$^{+13}_{-44}$ M$_{\odot}$), as with SN1983N (20$^{+6}_{-5}$ M$_{\odot}$).  Of the three unknowns, SN1945B did not have sufficient local stellar photometry for the fit to converge.  

SN1950B and SN1957D are candidates for CCSNe progenitors, with
  masses of 7.6$^{+3.3}_{-0.3}$ M$_{\odot}$ and 7.6$^{+8.2}_{-0.3}$
M$_{\odot}$ respectively.  This measurement for SN1957D is lower than
that of \citet{long2012}, who found a lower mass limit of $>$17
M$_{\odot}$, also looking at the local stellar population.  Our
analysis detects the presence of two populations at this location: one
at an age consistent with the analysis of \citet{long2012}, and an
underlying one at $\sim$40-50 Myr, which could have produced a
lower-mass progenitor.  Thus, this SNR is a good example of why it is
important to consider the full probability distribution, as the single
value with uncertainties does not always capture multiple peaks in the
age distribution.  Interestingly, a similar situation occurs with the
SNR 238, which \citet{blair2015} found to be hosted by a population of
very young stars, suggesting a progenitor with a mass $>$17
M$_{\odot}$.  We also detect this young population; however, we also
detect a population with an age of 40-50 Myr, making our mass
measurement 8.1$^{+12.5}_{-0.4}$ M$_{\odot}$.  Again, the
double-peaked age distribution is important to consider.

Thus, it appears that 5-6 of the 6 historical SNe are core-collapse (with SN1968L very likely core-collapse and SN1945B still unknown), which is similar to the fraction of the total SNR population and roughly consistent with expectations for a Sbc galaxy like M83 \citep[$\sim$20\%,][]{li2011}.  Furthermore, the 4 historical SNe for which we have measured ages appear to cover the full range of progenitor masses to which we are sensitive.

\section{Summary}

We measured resolved stellar photometry of seven archival \textit{HST} fields with almost complete coverage of M83.  We fit the photometry of stars within a 50~pc of 237 SNRs to derive the age distribution of the populations over the past 50~Myrs.  From these age distributions, we inferred probability distributions of 199 of the SNRs' progenitor masses.  The other 38 SNRs are good Type Ia candidates as they show no evidence for association with a young population in excess of the general field.  The spatial distribution of the 199 progenitor masses show that the most massive progenitors follow the spiral arms and inner disk.
    
The resulting mass distribution suggests that in M83, as in M31 and M33, the masses are dominated by progenitors of $<$20~M$_{\odot}$.  There are fewer progenitors with very young ages and high masses than expected from a standard IMF.  However, the measurement of some masses above 30~M$_{\odot}$, suggests that the low numbers are not due to a high end cutoff in the progenitor mass function.  Therefore, they may be due to the difficulty of finding SNRs and measuring resolved stellar photometry in the most dusty and dense star forming regions.   
    
While looking at the overall progenitor mass distributions measured here by using median masses and uncertainties and comparing to simple power-law distributions is of immediate interest for checking consistency with previous work and standard IMFs, we are also performing much more sophisticated statistical fitting of the full probability distributions of the masses (J. Murphy et al. in preparation).  By fitting the full probability distributions of the progenitor masses (Table~\ref{tab:probDistro}) with a reliable likelihood function we will determine the formal constraints on the intrinsic progenitor mass distribution.

Support for this work was provided by NASA through the grant AR-14325 from the Space Telescope Science Institute, which is operated by the Association of Universities for Research in Astronomy, Incorporated, under NASA contract NAS5-26555. Portions of this support were provided to University of Washington (BFW et al.), Johns Hopkins University (WPB), and Eureka Scientific (KSL). 

\software{Astrodrizzle \citep{hack2012}, DOLPHOT \citep{dolphin2000,dolphin2016}, PyRAF \citep{pyraf}, TinyTim PSFs \citep{krist2011}, MATCH \citep[including hybridMC;][]{dolphin2002,dolphin2012,dolphin2013}}


\startlongtable
\begin{deluxetable*}{cccccccccc}
	 \tablewidth{0pt}
     \tablecaption{M83 Supernova Remnants\label{tab:catalog}}
     \tablehead{
     	\colhead{Source ID} & \colhead{B12} & \colhead{B14} & \colhead{D10} & \colhead{L14} & \colhead{RA} & \colhead{DEC} & \colhead{Quality Flag} & \colhead{Use Flag}
     }
     \startdata
001 & B12-001 & -- & -- & - & 204.166625 & -29.859764 & 1b & o\\
002 & B12-002 & -- & -- & - & 204.168112 & -29.851814 & 2 & o\\
003 & B12-003 & -- & -- & X019 & 204.1704 & -29.854906 & 1a & o\\
004 & B12-004 & -- & -- & - & 204.172921 & -29.871072 & 1c & o\\
005 & B12-005 & -- & -- & - & 204.173246 & -29.8323 & 1b & o\\
006 & B12-006 & -- & -- & - & 204.176375 & -29.871492 & 1c & o\\
007 & B12-007 & -- & -- & - & 204.178012 & -29.876375 & 1c & o\\
008 & B12-008 & -- & -- & - & 204.182079 & -29.846078 & 1c & o\\
009 & B12-009 & -- & -- & - & 204.182579 & -29.869775 & 1c & o\\
010 & B12-010 & -- & -- & - & 204.186 & -29.842747 & 1b & o\\
011 & B12-011 & -- & -- & - & 204.1888 & -29.885486 & 1c & o\\
012 & B12-012 & -- & -- & - & 204.190258 & -29.872578 & 1b & o\\
013 & B12-013 & -- & -- & - & 204.191375 & -29.892878 & 1c & o\\
014 & B12-014 & -- & -- & - & 204.1934 & -29.895092 & 2 & o\\
015 & B12-015 & -- & -- & - & 204.195592 & -29.778244 & 2 & o\\
016 & B12-016 & -- & -- & - & 204.196371 & -29.925422 & 2 & o\\
017 & B12-017 & -- & -- & - & 204.196583 & -29.897611 & 2 & o\\
018 & B12-018 & -- & -- & - & 204.196758 & -29.893589 & 1c & o\\
019 & B12-019 & -- & -- & - & 204.197083 & -29.817708 & 1c & o\\
020 & B12-020 & -- & -- & - & 204.199279 & -29.855039 & 1b & i\\
021 & B12-021 & -- & -- & - & 204.199721 & -29.862778 & 2 & i\\
022 & B12-307 & -- & -- & - & 204.199958 & -29.890722 & 2 & o\\
023 & B12-022 & -- & -- & - & 204.200446 & -29.859408 & 1b & n\\
024 & B12-023 & -- & -- & X046 & 204.201233 & -29.879075 & 1a & p\\
025 & B12-024 & -- & -- & - & 204.201883 & -29.861719 & 1c & n\\
026 & B12-025 & -- & -- & - & 204.20245 & -29.868144 & 1b & n\\
027 & B12-026 & B14-01 & -- & - & 204.204138 & -29.881692 & 1b & p\\
028 & B12-027 & -- & -- & - & 204.204688 & -29.873583 & 1c & p\\
029 & B12-028 & -- & -- & - & 204.205692 & -29.888867 & 2 & p\\
030 & B12-029 & -- & -- & - & 204.206408 & -29.8603 & 2 & p\\
031 & B12-030 & -- & -- & - & 204.206687 & -29.884908 & 1c & p\\
032 & B12-031 & -- & -- & - & 204.206762 & -29.887125 & 1b & p\\
033 & B12-32 & -- & -- & - & 204.2068 & -29.8429 & 2 & p\\
034 & B12-033 & -- & -- & - & 204.207004 & -29.901153 & 1b & p\\
035 & B12-034 & -- & -- & - & 204.207158 & -29.849208 & 1b & p\\
036 & B12-036 & B14-02 & -- & X053 & 204.207538 & -29.871375 & 1a & p\\
037 & B12-035 & -- & -- & - & 204.207546 & -29.885639 & 1b & n\\
038 & B12-309 & -- & -- & - & 204.207896 & -29.883083 & 2 & n\\
039 & -- & B14-03 & -- & - & 204.208817 & -29.878797 & 1g & p\\
040 & B12-037 & B14-04 & -- & X057 & 204.208821 & -29.88575 & 1a & p\\
041 & B12-038 & -- & -- & - & 204.209292 & -29.856747 & 1c & i\\
042 & B12-310 & -- & -- & - & 204.209333 & -29.843581 & 2 & p\\
043 & B12-039 & -- & -- & - & 204.209517 & -29.879861 & 1b & p\\
044 & B12-040 & -- & -- & - & 204.210408 & -29.863831 & 1c & p\\
045 & B12-041 & B14-05 & -- & X061 & 204.210633 & -29.884411 & 1d & p\\
046 & B12-042 & -- & -- & - & 204.211204 & -29.878206 & 2 & p\\
047 & B12-043 & -- & -- & - & 204.211492 & -29.886275 & 1b & p\\
048 & B12-044 & -- & -- & - & 204.211704 & -29.838556 & 2 & p\\
049 & B12-045 & -- & -- & X063 & 204.211896 & -29.877658 & 1a & p\\
SN1983N & -- & -- & -- & - & 204.21204 & -29.901278 & h & p\\
051 & B12-047 & -- & -- & X064 & 204.212154 & -29.882981 & 1d & p\\
052 & B12-046 & -- & -- & - & 204.212167 & -29.867728 & 1c & p\\
053 & B12-048 & -- & -- & X065 & 204.212475 & -29.873872 & 1a & p\\
054 & B12-049 & B14-06 & -- & - & 204.212579 & -29.883711 & 2 & p\\
055 & -- & B14-07 & -- & X067 & 204.2133 & -29.845089 & 1e & p\\
056 & B12-050 & -- & -- & - & 204.213458 & -29.877964 & 2 & p\\
057 & B12-051 & -- & -- & - & 204.213825 & -29.835297 & 2 & p\\
058 & B12-052 & -- & -- & - & 204.214292 & -29.824497 & 1c & p\\
059 & -- & B14-08 & -- & - & 204.214512 & -29.8759 & 1g & p\\
060 & -- & B14-09 & -- & - & 204.214692 & -29.883589 & 2 & p\\
061 & B12-053 & -- & -- & - & 204.214942 & -29.880575 & 1c & p\\
062 & B12-054 & -- & -- & - & 204.215421 & -29.874344 & 2 & p\\
063 & -- & B14-10 & -- & - & 204.215858 & -29.867194 & 1g & p\\
064 & B12-55 & -- & -- & - & 204.216562 & -29.914561 & 1c & p\\
065 & B12-311 & -- & -- & - & 204.217796 & -29.905803 & 2 & n\\
066 & B12-056 & -- & -- & - & 204.218092 & -29.842544 & 1c & n\\
067 & B12-058 & -- & -- & - & 204.218258 & -29.868097 & 2 & n\\
068 & B12-057 & -- & -- & - & 204.218321 & -29.845536 & 1b & p\\
069 & B12-059 & -- & -- & - & 204.2188 & -29.825731 & 2 & p\\
070 & B12-060 & -- & -- & X093 & 204.219329 & -29.8782 & 1d & p\\
071 & B12-061 & -- & -- & - & 204.219904 & -29.857919 & 2 & p\\
SN1950B & -- & -- & -- & - & 204.22071 & -29.865861 & h & p\\
073 & B12-062 & -- & -- & - & 204.220771 & -29.839964 & 1c & i\\
074 & B12-063 & -- & -- & - & 204.221183 & -29.871217 & 1d & p\\
075 & B12-064 & -- & -- & - & 204.221587 & -29.874806 & 2 & n\\
076 & B12-065 & B14-11 & -- & X105 & 204.221783 & -29.890369 & 1a & p\\
077 & B12-067 & B14-12 & -- & X106 & 204.222046 & -29.88005 & 1a & p\\
078 & B12-066 & -- & -- & X107 & 204.222062 & -29.878478 & 1a & n\\
079 & B12-068 & -- & -- & - & 204.222179 & -29.930956 & 2 & o\\
080 & B12-069 & -- & -- & - & 204.222367 & -29.843989 & 1b & n\\
081 & B12-070 & -- & -- & - & 204.222967 & -29.877269 & 1c & n\\
SN1945B & -- & -- & -- & - & 204.223335 & -29.915556 & h & i\\
083 & B12-312 & -- & -- & X110 & 204.223337 & -29.933567 & 1e & o\\
084 & B12-071 & -- & -- & - & 204.223446 & -29.879417 & 1c & n\\
085 & -- & B14-13 & -- & - & 204.223871 & -29.814239 & 2 & p\\
086 & B12-072 & -- & -- & - & 204.224054 & -29.91135 & 1c & n\\
087 & B12-073 & -- & -- & X116 & 204.224558 & -29.813383 & 1a & p\\
088 & B12-074 & -- & -- & X119 & 204.225671 & -29.86925 & 1a & p\\
089 & B12-075 & B14-14 & -- & X121 & 204.226012 & -29.841156 & 1d & p\\
090 & B12-076 & -- & -- & - & 204.226433 & -29.838208 & 1c & p\\
091 & B12-077 & -- & -- & - & 204.226858 & -29.933436 & 1f & o\\
092 & B12-078 & -- & -- & - & 204.226937 & -29.848047 & 2 & p\\
093 & B12-079 & -- & -- & - & 204.227054 & -29.84065 & 1c & p\\
094 & B12-080 & -- & -- & - & 204.227608 & -29.884706 & 2 & p\\
095 & B12-081 & -- & -- & - & 204.227646 & -29.883658 & 2 & p\\
096 & B12-082 & -- & -- & X127 & 204.2283 & -29.883167 & 1d & p\\
097 & B12-083 & -- & -- & - & 204.228442 & -29.884647 & 2 & p\\
098 & B12-085 & -- & -- & - & 204.228521 & -29.831553 & 2 & p\\
099 & B12-084 & -- & -- & X128 & 204.228617 & -29.838492 & 1a & p\\
100 & B12-086 & -- & -- & - & 204.228917 & -29.796075 & 2 & o\\
101 & B12-088 & -- & -- & - & 204.229304 & -29.856922 & 1c & p\\
102 & B12-087 & -- & -- & X129 & 204.229312 & -29.877658 & 1a & n\\
103 & B12-313 & -- & -- & - & 204.229371 & -29.915103 & 2 & p\\
104 & B12-089 & -- & -- & X131 & 204.229437 & -29.884578 & 1a & p\\
105 & B12-090 & -- & -- & X134 & 204.229692 & -29.844531 & 1d & p\\
106 & B12-091 & -- & -- & - & 204.230075 & -29.884717 & 1b & p\\
107 & B12-314 & B14-15 & -- & - & 204.230312 & -29.900789 & 2 & p\\
108 & B12-092 & -- & -- & - & 204.230471 & -29.843678 & 1c & p\\
109 & B12-094 & -- & -- & - & 204.230513 & -29.832406 & 2 & p\\
110 & B12-093 & -- & -- & X136 & 204.230638 & -29.848253 & 1d & p\\
111 & B12-95 & -- & -- & - & 204.230858 & -29.810886 & 1c & p\\
112 & B12-096 & -- & -- & - & 204.231125 & -29.884297 & 1c & p\\
113 & B12-097 & -- & -- & - & 204.231146 & -29.878808 & 1b & p\\
114 & B12-098 & -- & -- & - & 204.231729 & -29.884336 & 1b & p\\
115 & B12-099 & -- & -- & - & 204.231729 & -29.793711 & 2 & o\\
116 & B12-100 & -- & -- & - & 204.232054 & -29.823644 & 2 & p\\
117 & B12-101 & -- & -- & - & 204.232492 & -29.855483 & 1b & p\\
118 & B12-102 & -- & -- & - & 204.232625 & -29.885894 & 2 & p\\
119 & B12-103 & -- & -- & - & 204.232967 & -29.886375 & 1c & p\\
120 & B12-104 & -- & -- & - & 204.2336 & -29.934903 & 1b & o\\
121 & B12-105 & -- & -- & - & 204.233779 & -29.826392 & 2 & p\\
122 & B12-106 & B14-16 & -- & X141 & 204.234279 & -29.881994 & 1a & n\\
123 & B12-107 & -- & -- & - & 204.234479 & -29.887067 & 1c & p\\
124 & B12-108 & -- & -- & - & 204.234842 & -29.825644 & 2 & p\\
125 & B12-109 & B14-17 & -- & X149 & 204.2367 & -29.830461 & 1a & p\\
126 & B12-110 & -- & -- & - & 204.236742 & -29.823558 & 1b & p\\
127 & B12-111 & -- & -- & - & 204.237183 & -29.902956 & 1c & n\\
128 & B12-112 & -- & -- & - & 204.238154 & -29.892706 & 1b & p\\
129 & B12-113 & -- & -- & - & 204.240975 & -29.801625 & 2 & p\\
130 & B12-115 & B14-18 & -- & X159 & 204.241167 & -29.884097 & 1a & p\\
131 & B12-114 & -- & -- & - & 204.241525 & -29.803514 & 1c & p\\
132 & B12-316 & -- & -- & - & 204.241833 & -29.817222 & 2 & o\\
133 & B12-116 & -- & -- & - & 204.241975 & -29.8958 & 1c & p\\
134 & B12-117 & -- & -- & X166 & 204.243971 & -29.805475 & 1e & p\\
135 & -- & B14-19 & -- & - & 204.244346 & -29.851803 & 1b & p\\
136 & B12-118 & -- & D10-01 & X172 & 204.244683 & -29.850156 & 1a & n\\
137 & -- & B14-20 & D10-02 & - & 204.245412 & -29.873961 & 1b & p\\
138 & B12-119 & B14-21 & -- & - & 204.245837 & -29.882442 & 2 & p\\
139 & B12-120 & -- & -- & - & 204.245854 & -29.883708 & 1c & p\\
140 & B12-318 & -- & -- & - & 204.24595 & -29.916283 & 2 & i\\
141 & B12-121 & -- & -- & - & 204.246271 & -29.895386 & 2 & p\\
142 & -- & B14-22 & D10-03 & X181 & 204.246537 & -29.863306 & 1d & n\\
143 & B12-319 & -- & -- & - & 204.247079 & -29.916158 & 2 & p\\
144 & -- & B14-23 & -- & - & 204.24715 & -29.810142 & 2 & p\\
145 & B12-122 & -- & -- & X183 & 204.247217 & -29.919153 & 1a & p\\
146 & B12-123 & -- & -- & X184 & 204.247262 & -29.810469 & 1d & p\\
147 & -- & B14-24 & -- & - & 204.247675 & -29.810275 & 2 & p\\
148 & B12-320 & -- & -- & - & 204.247687 & -29.909628 & 2 & n\\
149 & B12-125 & -- & -- & - & 204.247796 & -29.821292 & 1b & n\\
150 & B12-124 & -- & D10-04 & X186 & 204.247933 & -29.867761 & 1a & p\\
151 & B12-126 & -- & D10-05 & - & 204.248683 & -29.842431 & 1c & n\\
152 & -- & B14-25 & -- & - & 204.249121 & -29.810519 & 2 & p\\
153 & B12-127 & -- & -- & X195 & 204.249371 & -29.923881 & 1a & i\\
154 & B12-128 & -- & -- & - & 204.250042 & -29.809369 & 1c & n\\
155 & B12-129 & B14-26 & -- & X199 & 204.250137 & -29.904708 & 1a & n\\
156 & -- & B14-27 & D10-N-01 & X202 & 204.250188 & -29.867206 & 1d & c\\
157 & -- & B14-28 & D10-06 & - & 204.250271 & -29.869097 & 1c & c\\
158 & B12-130 & -- & -- & - & 204.250354 & -29.811194 & 2 & p\\
159 & B12-131 & -- & -- & X205 & 204.250654 & -29.802817 & 1d & p\\
160 & -- & -- & D10-N-02 & - & 204.250883 & -29.866214 & 1h & c\\
161 & B12-132 & -- & D10-07 & - & 204.2514 & -29.855756 & 1b & n\\
162 & -- & -- & D10-N-05 & - & 204.251608 & -29.866303 & 1c & c\\
SN1968L & -- & -- & -- & - & 204.251625 & -29.866250 & h & c\\
164 & -- & -- & D10-N-06 & - & 204.251654 & -29.867142 & 1h & c\\
165 & B12-133 & -- & -- & X215 & 204.251654 & -29.889697 & 1a & p\\
166 & -- & B14-30 & D10-N-07 & - & 204.251725 & -29.868392 & 1h & c\\
167 & -- & B14-31 & -- & - & 204.251729 & -29.872931 & 1c & p\\
168 & -- & B14-32 & -- & - & 204.252292 & -29.868486 & 2 & c\\
169 & -- & -- & D10-N-08 & - & 204.252308 & -29.866319 & 1c & c\\
170 & -- & -- & D10-N-09 & - & 204.252496 & -29.869111 & 1h & c\\
171 & -- & -- & D10-N-10 & - & 204.252692 & -29.866497 & 1h & c\\
172 & B12-134 & -- & -- & - & 204.252817 & -29.907414 & 1b & p\\
173 & -- & -- & D10-N-11 & - & 204.252825 & -29.865856 & 1h & c\\
174 & B12-135 & -- & D10-08 & - & 204.252854 & -29.872781 & 2 & p\\
175 & -- & -- & D10-N-12 & - & 204.252938 & -29.866636 & 1h & c\\
176 & B12-136 & -- & -- & - & 204.253062 & -29.889928 & 2 & p\\
177 & -- & -- & D10-N-13 & - & 204.253171 & -29.868306 & 1h & c\\
178 & -- & B14-34 & -- & - & 204.25365 & -29.869014 & 2 & c\\
179 & -- & -- & D10-N-14 & - & 204.253892 & -29.865042 & 1h & c\\
180 & -- & -- & D10-N-14 & - & 204.253896 & -29.865181 & 1g & c\\
181 & -- & -- & D10-N-15 & - & 204.253933 & -29.865522 & 1h & c\\
182 & -- & -- & D10-N-14 & - & 204.254054 & -29.864886 & 1g & c\\
183 & B12-137 & B14-35 & D10-09 & X235 & 204.254238 & -29.848983 & 1a & p\\
184 & B12-138 & -- & -- & - & 204.254425 & -29.904408 & 1c & p\\
185 & -- & B14-36 & D10-10 & - & 204.254463 & -29.861544 & 2 & p\\
186 & -- & -- & D10-N-16 & - & 204.254646 & -29.86445 & 1z & c\\
187 & B12-139 & -- & -- & - & 204.254817 & -29.952981 & 1c & o\\
188 & -- & -- & D10-N-17 & X241 & 204.254892 & -29.865928 & 1d & c\\
189 & B12-321 & B14-38 & -- & X243 & 204.255313 & -29.866636 & 1z & c\\
190 & -- & -- & D10-N-18 & - & 204.255479 & -29.865956 & 1h & c\\
191 & B12-140 & -- & -- & - & 204.256304 & -29.837425 & 1c & n\\
192 & B12-141 & -- & -- & X249 & 204.25645 & -29.833019 & 1d & p\\
193 & -- & B14-39 & D10-N-19 & X250 & 204.256717 & -29.867206 & 1d & c\\
194 & B12-142 & -- & -- & X253 & 204.256933 & -29.902803 & 1a & p\\
195 & B12-143 & -- & D10-12 & X256 & 204.257133 & -29.853711 & 1d & p\\
196 & B12-144 & -- & -- & X255 & 204.257167 & -29.911217 & 1d & p\\
197 & -- & B14-37 & -- & - & 204.258308 & -29.864289 & 1g & p\\
198 & B12-145 & -- & D10-13 & - & 204.258492 & -29.880444 & 1c & p\\
199 & B12-146 & -- & D10-14 & - & 204.258612 & -29.866197 & 1b & p\\
200 & B12-147 & B14-40 & -- & X261 & 204.2592 & -29.831231 & 1a & i\\
201 & B12-148 & -- & -- & X262 & 204.25965 & -29.835272 & 1a & p\\
202 & B12-149 & -- & -- & - & 204.260058 & -29.90915 & 1c & p\\
203 & B12-150 & B14-41 & D10-15 & X265 & 204.260096 & -29.857236 & 1a & p\\
204 & -- & B14-42 & -- & - & 204.262033 & -29.810861 & 2 & p\\
205 & B12-151 & B14-43 & -- & X272 & 204.262563 & -29.829294 & 1e & p\\
206 & -- & B14-44 & -- & - & 204.263096 & -29.9047 & 1g & n\\
207 & -- & B14-45 & -- & - & 204.264183 & -29.900689 & 1g & p\\
208 & B12-152 & -- & -- & - & 204.264512 & -29.846303 & 2 & p\\
SN1957D & B12-324 & B14-46 & -- & X279 & 204.264917 & -29.827981 & h & p\\
210 & B12-153 & -- & -- & - & 204.266167 & -29.828614 & 2 & p\\
211 & B12-154 & -- & -- & - & 204.2669 & -29.900539 & 1b & p\\
212 & -- & -- & D10-1-01 & - & 204.267158 & -29.851053 & 2 & n\\
213 & B12-155 & -- & -- & - & 204.267242 & -29.887847 & 1b & p\\
214 & B12-156 & -- & -- & X287 & 204.268383 & -29.827403 & 1a & p\\
215 & B12-158 & -- & -- & - & 204.268533 & -29.900939 & 2 & p\\
216 & B12-157 & -- & -- & - & 204.268596 & -29.896589 & 1b & p\\
217 & B12-159 & -- & -- & X288 & 204.26875 & -29.826542 & 1a & p\\
218 & B12-160 & -- & -- & X292 & 204.269638 & -29.926342 & 1a & n\\
219 & B12-161 & -- & -- & - & 204.269833 & -29.898228 & 2 & p\\
220 & B12-162 & -- & -- & - & 204.270054 & -29.835219 & 1b & p\\
221 & B12-163 & -- & -- & - & 204.2702 & -29.828344 & 2 & n\\
222 & -- & -- & D10-17 & - & 204.270321 & -29.871828 & 1b & n\\
223 & B12-164 & -- & -- & - & 204.270683 & -29.837872 & 1c & n\\
224 & -- & B14-47 & -- & - & 204.270933 & -29.922703 & 1g & p\\
225 & B12-165 & -- & -- & - & 204.273271 & -29.915633 & 1c & p\\
226 & B12-166 & -- & D10-18 & - & 204.274162 & -29.879456 & 1c & n\\
227 & B12-167 & -- & -- & - & 204.274462 & -29.917772 & 2 & p\\
228 & -- & B14-49 & -- & - & 204.2745 & -29.845961 & 2 & p\\
229 & B12-327 & -- & -- & - & 204.274504 & -29.819794 & 2 & p\\
230 & B12-168 & -- & -- & - & 204.275004 & -29.834508 & 1b & p\\
231 & B12-169 & -- & -- & X310 & 204.275146 & -29.920647 & 1a & p\\
232 & B12-170 & -- & -- & X311 & 204.275671 & -29.912089 & 1a & i\\
233 & -- & B14-50 & -- & - & 204.275971 & -29.918081 & 2 & p\\
234 & B12-171 & -- & D10-19 & X313 & 204.276821 & -29.840269 & 1a & p\\
235 & B12-172 & -- & -- & - & 204.276854 & -29.907578 & 1b & p\\
236 & B12-173 & -- & -- & - & 204.276854 & -29.835061 & 1c & p\\
237 & B12-174 & -- & -- & - & 204.277688 & -29.892786 & 1c & p\\
238 & B12-174a & -- & -- & X316 & 204.277692 & -29.892392 & 1z & p\\
239 & B12-175 & -- & -- & - & 204.278433 & -29.823953 & 2 & p\\
240 & -- & -- & D10-1-02 & - & 204.279042 & -29.849206 & 2 & p\\
241 & B12-328 & -- & -- & - & 204.279046 & -29.915914 & 2 & p\\
242 & -- & B14-51 & D10-20 & - & 204.279121 & -29.852664 & 2 & p\\
243 & B12-177 & -- & -- & X319 & 204.279142 & -29.818856 & 1d & p\\
244 & B12-176 & -- & -- & - & 204.279171 & -29.904494 & 1c & n\\
245 & B12-178 & -- & -- & X320 & 204.279496 & -29.889158 & 1a & p\\
246 & B12-329 & -- & -- & - & 204.279546 & -29.820414 & 2 & p\\
247 & B12-179 & B14-52 & -- & - & 204.279604 & -29.850431 & 1b & p\\
248 & B12-180 & -- & D10-22 & X326 & 204.281129 & -29.859267 & 1a & p\\
249 & B12-181 & -- & -- & - & 204.28125 & -29.904494 & 1b & n\\
250 & B12-182 & -- & D10-23 & - & 204.281538 & -29.872003 & 1c & n\\
251 & B12-183 & B14-53 & D10-21 & - & 204.282038 & -29.852792 & 1f & p\\
252 & B12-184 & -- & -- & - & 204.282113 & -29.883692 & 1b & p\\
253 & B12-185 & -- & -- & - & 204.282488 & -29.9034 & 1b & p\\
254 & B12-186 & -- & -- & X330 & 204.283 & -29.822253 & 1d & p\\
255 & B12-187 & -- & -- & - & 204.283375 & -29.854592 & 2 & p\\
256 & B12-188 & -- & D10-25 & - & 204.283746 & -29.872636 & 1b & p\\
257 & B12-189 & -- & -- & - & 204.283983 & -29.889078 & 2 & p\\
258 & -- & B14-54 & D10-26 & X336 & 204.284713 & -29.848981 & 1a & p\\
259 & -- & B14-55 & -- & - & 204.284983 & -29.879919 & 2 & p\\
260 & -- & B14-56 & -- & - & 204.285096 & -29.869508 & 2 & p\\
261 & B12-190 & -- & D10-27 & - & 204.285333 & -29.867222 & 1d & p\\
262 & B12-191 & -- & D10-28 & X339 & 204.285696 & -29.859772 & 1a & p\\
263 & B12-333 & B14-57 & D10-29 & X341 & 204.285983 & -29.878556 & 1e & p\\
264 & B12-192 & -- & D10-30 & - & 204.286025 & -29.864828 & 1c & p\\
265 & B12-193 & -- & D10-32 & X342 & 204.286496 & -29.860425 & 1a & p\\
266 & B12-194 & -- & D10-33 & - & 204.287708 & -29.859261 & 1b & p\\
SN1923A & -- & -- & -- & - & 204.28827 & -29.850386 & h & p\\
268 & B12-195 & -- & D10-34 & X348 & 204.288446 & -29.859433 & 1d & p\\
269 & -- & B14-58 & D10-35 & - & 204.288813 & -29.849592 & 2 & p\\
270 & B12-196 & -- & -- & - & 204.290363 & -29.891717 & 1c & p\\
271 & B12-197 & -- & D10-36 & X350 & 204.291992 & -29.857828 & 1a & p\\
272 & B12-198 & -- & -- & - & 204.292454 & -29.838372 & 2 & p\\
273 & B12-334 & -- & -- & - & 204.292475 & -29.816472 & 2 & p\\
274 & B12-199 & -- & D10-37 & X352 & 204.292963 & -29.858058 & 1a & p\\
275 & -- & -- & D10-38 & - & 204.293183 & -29.859425 & 1b & p\\
276 & B12-200 & -- & -- & - & 204.294921 & -29.832372 & 2 & p\\
277 & B12-201 & -- & D10-39 & - & 204.294996 & -29.862403 & 1b & p\\
278 & -- & -- & D10-40 & - & 204.295163 & -29.879019 & 1b & p\\
279 & B12-202 & -- & -- & - & 204.295525 & -29.831381 & 1b & p\\
280 & B12-203 & -- & -- & X353 & 204.295708 & -29.8462 & 1d & p\\
281 & B12-204 & -- & -- & - & 204.296283 & -29.888086 & 1b & n\\
282 & B12-205 & -- & -- & - & 204.297229 & -29.905403 & 2 & p\\
283 & B12-207 & -- & -- & X355 & 204.297771 & -29.837117 & 1a & p\\
284 & B12-206 & -- & -- & X356 & 204.297808 & -29.861489 & 1a & p\\
285 & B12-208 & -- & -- & - & 204.29845 & -29.860981 & 2 & p\\
286 & B12-209 & -- & -- & - & 204.299483 & -29.871036 & 1a & p\\
287 & B12-336 & B14-59 & -- & X360 & 204.30035 & -29.849208 & 1e & p\\
288 & B12-210 & -- & -- & X364 & 204.3019 & -29.838903 & 1a & o\\
289 & B12-211 & -- & -- & X368 & 204.303367 & -29.836678 & 1a & o\\
290 & B12-338 & -- & -- & - & 204.303392 & -29.9123 & 2 & i\\
291 & B12-212 & -- & -- & - & 204.303517 & -29.910761 & 1c & n\\
292 & -- & B14-60 & -- & - & 204.304442 & -29.860683 & 2 & p\\
293 & B12-213 & -- & -- & - & 204.304488 & -29.855053 & 1b & p\\
294 & -- & B14-61 & -- & - & 204.305525 & -29.859972 & 2 & p\\
295 & B12-215 & -- & -- & - & 204.308208 & -29.864208 & 1b & p\\
296 & B12-214 & -- & -- & - & 204.308442 & -29.881753 & 1c & p\\
297 & B12-216 & -- & -- & - & 204.309812 & -29.8351 & 1c & o\\
298 & B12-217 & -- & -- & - & 204.310125 & -29.839233 & 1c & p\\
299 & B12-218 & -- & -- & - & 204.311167 & -29.842789 & 2 & p\\
300 & B12-219 & -- & -- & - & 204.311829 & -29.916283 & 1b & o\\
301 & B12-220 & -- & -- & - & 204.316771 & -29.884439 & 1b & o\\
302 & B12-221 & B14-62 & -- & X389 & 204.321675 & -29.864825 & 1a & p\\
303 & B12-222 & -- & -- & - & 204.321942 & -29.890267 & 1b & o\\
304 & B12-223 & B14-63 & -- & X391 & 204.322612 & -29.864969 & 1a & p\\
305 & B12-224 & -- & -- & - & 204.322892 & -29.893278 & 1c & o\\
306 & B12-344 & -- & -- & - & 204.324125 & -29.865397 & 2 & p\\
307 & B12-225 & -- & -- & - & 204.328079 & -29.897375 & 1c & o\\
     \enddata
\end{deluxetable*}

\startlongtable
\begin{deluxetable*}{cccccccc}
	 \tablewidth{0pt}
     \tablecaption{SNR photometry depth, extinction, and progenitor mass estimates\label{tab:report}}
     \tablehead{
     	\colhead{} \vspace{-0.1cm} & \multicolumn{3}{c}{50\% Completeness Limit}\\ \cline{2-4}
     	\colhead{ID} & \colhead{F336W} & \colhead{F438W} \vspace{-0.3cm}& \colhead{F814W} & \colhead{Count$_{F438W}$} & \colhead{dAv} & \colhead{Av} & \colhead{$M_{ZAMS}$}\\
     }
     \startdata
024 & 26.0 & 27.0 & 25.9 & 425 & 0.0 & 0.2 & $14.86^{+0.84}_{-1.46}$ \\
027 & 26.2 & 27.2 & 26.1 & 173 & 0.0 & 1.6 & $8.43^{+1.87}_{-0.73}$ \\
028 & 26.0 & 27.0 & 25.8 & 64 & 0.0 & 1.8 & $11.37^{+4.33}_{-4.07}$ \\
029 & 26.3 & 27.1 & 26.1 & 150 & 0.0 & 0.2 & $8.07^{+4.43}_{-0.37}$ \\
030 & 26.1 & 27.0 & 26.0 & 98 & 0.4 & 0.8 & $7.64^{+5.76}_{-0.34}$ \\
031 & 25.9 & 26.8 & 25.3 & 57 & 0.4 & 1.2 & $9.94^{+1.76}_{-2.24}$ \\
032 & 26.3 & 27.1 & 26.0 & 187 & 0.0 & 0.2 & $9.90^{+7.20}_{-1.30}$ \\
033 & 26.3 & 27.1 & 26.1 & 149 & 0.0 & 0.2 & $9.93^{+3.47}_{-0.93}$ \\
034 & 25.9 & 26.8 & 25.3 & 47 & 0.0 & 1.2 & $43.16^{+8.94}_{-34.16}$ \\
035 & 26.0 & 27.0 & 25.8 & 52 & 0.0 & 0.4 & $27.14^{+2.06}_{-17.54}$ \\
036 & 26.3 & 27.1 & 26.0 & 177 & 0.0 & 0.6 & $9.50^{+3.90}_{-0.50}$ \\
039 & 26.1 & 27.0 & 25.9 & 203 & 1.6 & 0.8 & $12.17^{+4.93}_{-2.57}$ \\
040 & 26.0 & 26.7 & 25.1 & 40 & 0.0 & 2.4 & $15.10^{+8.00}_{-7.80}$ \\
042 & 26.3 & 27.0 & 26.0 & 190 & 0.6 & 0.6 & $9.56^{+2.14}_{-0.56}$ \\
043 & 26.1 & 27.0 & 25.9 & 222 & 0.0 & 0.2 & $10.86^{+0.84}_{-0.56}$ \\
044 & 26.0 & 26.9 & 25.5 & 105 & 0.0 & 0.2 & $9.73^{+3.67}_{-0.73}$ \\
045 & 26.0 & 26.9 & 25.5 & 125 & 0.2 & 0.8 & $10.65^{+1.85}_{-1.65}$ \\
046 & 26.2 & 26.9 & 26.0 & 161 & 0.6 & 1.0 & $16.78^{+1.92}_{-5.88}$ \\
047 & 26.0 & 27.1 & 25.8 & 86 & 0.0 & 1.2 & $8.80^{+5.70}_{-0.70}$ \\
048 & 26.0 & 27.0 & 25.8 & 54 & 0.0 & 0.4 & $8.53^{+3.97}_{-0.43}$ \\
049 & 26.3 & 27.3 & 26.3 & 136 & 0.4 & 1.0 & $11.29^{+2.11}_{-0.99}$ \\
051 & 26.3 & 27.1 & 26.1 & 162 & 0.8 & 1.0 & $8.98^{+3.52}_{-0.38}$ \\
052 & 25.8 & 26.5 & 25.2 & 770 & 0.6 & 0.2 & $10.68^{+5.02}_{-0.38}$ \\
053 & 25.5 & 26.4 & 25.5 & 894 & 2.2 & 0.2 & $18.40^{+0.30}_{-6.70}$ \\
054 & 26.3 & 27.1 & 26.1 & 175 & 0.0 & 0.6 & $11.17^{+3.33}_{-2.57}$ \\
055 & 26.3 & 27.1 & 26.0 & 120 & 0.6 & 1.0 & $21.22^{+1.88}_{-11.62}$ \\
056 & 26.1 & 27.0 & 25.9 & 183 & 1.0 & 0.8 & $10.09^{+8.61}_{-0.49}$ \\
057 & 26.3 & 27.0 & 26.0 & 131 & 0.4 & 0.4 & $59.00^{+7.60}_{-38.40}$ \\
058 & 26.0 & 27.1 & 25.8 & 41 & 0.0 & 1.0 & $9.33^{+9.37}_{-0.73}$ \\
059 & 26.0 & 26.7 & 25.1 & 36 & 0.2 & 2.6 & $16.77^{+0.33}_{-7.77}$ \\
060 & 25.9 & 26.9 & 25.8 & 144 & 0.8 & 1.2 & $21.49^{+1.61}_{-13.39}$ \\
061 & 26.3 & 27.1 & 26.1 & 190 & 0.0 & 0.6 & $9.56^{+1.34}_{-0.56}$ \\
062 & 26.0 & 26.9 & 25.7 & 205 & 0.0 & 1.0 & $11.80^{+1.60}_{-2.80}$ \\
063 & 26.1 & 27.0 & 26.0 & 187 & 0.0 & 0.8 & $7.53^{+4.17}_{-0.23}$ \\
064 & 25.9 & 26.8 & 25.3 & 41 & 0.0 & 0.4 & $27.56^{+1.64}_{-15.86}$ \\
068 & 26.5 & 27.3 & 26.3 & 65 & 0.0 & 2.0 & $12.06^{+0.44}_{-4.76}$ \\
069 & 26.0 & 27.1 & 25.8 & 41 & 0.0 & 0.6 & $7.93^{+7.77}_{-0.23}$ \\
070 & 26.0 & 26.8 & 25.9 & 384 & 0.0 & 0.2 & $8.02^{+2.28}_{-0.32}$ \\
071 & 26.0 & 27.0 & 25.9 & 199 & 0.0 & 0.6 & $8.97^{+5.53}_{-0.37}$ \\
074 & 26.1 & 27.1 & 25.9 & 170 & 0.2 & 0.2 & $9.94^{+5.76}_{-0.34}$ \\
076 & 25.5 & 26.4 & 25.5 & 342 & 0.8 & 0.2 & $14.66^{+4.04}_{-3.76}$ \\
077 & 25.9 & 26.9 & 25.9 & 593 & 2.2 & 0.6 & $7.71^{+3.99}_{-0.41}$ \\
085 & 25.8 & 26.5 & 25.2 & 180 & 0.0 & 0.2 & $22.24^{+12.46}_{-11.94}$ \\
087 & 26.0 & 26.9 & 26.1 & 274 & 0.8 & 0.2 & $16.23^{+12.97}_{-6.63}$ \\
088 & 26.1 & 27.1 & 25.8 & 78 & 0.0 & 2.8 & $7.53^{+27.17}_{-0.23}$ \\
089 & 26.2 & 26.9 & 26.0 & 298 & 1.0 & 0.4 & $10.77^{+7.93}_{-0.47}$ \\
090 & 26.2 & 27.2 & 26.1 & 35 & 0.0 & 2.0 & $9.33^{+11.27}_{-0.73}$ \\
092 & 26.1 & 27.1 & 25.9 & 133 & 0.0 & 1.6 & $12.09^{+2.41}_{-4.39}$ \\
093 & 26.0 & 26.9 & 25.5 & 97 & 0.2 & 1.0 & $9.96^{+0.34}_{-0.36}$ \\
094 & 26.3 & 27.0 & 26.0 & 181 & 0.8 & 1.0 & $9.45^{+3.95}_{-0.45}$ \\
095 & 26.1 & 27.1 & 25.9 & 132 & 0.0 & 1.2 & $9.40^{+1.50}_{-0.80}$ \\
096 & 26.0 & 27.0 & 25.9 & 208 & 3.0 & 1.2 & $10.92^{+0.78}_{-2.82}$ \\
097 & 26.1 & 27.1 & 25.9 & 220 & 2.0 & 0.2 & $11.03^{+0.67}_{-1.43}$ \\
098 & 26.3 & 27.3 & 26.3 & 153 & 0.0 & 0.2 & $7.55^{+4.95}_{-0.25}$ \\
099 & 26.5 & 27.3 & 26.3 & 78 & 0.0 & 2.0 & $7.93^{+15.17}_{-0.63}$ \\
101 & 26.0 & 27.1 & 25.8 & 82 & 0.0 & 1.6 & $15.06^{+0.64}_{-7.76}$ \\
103 & 25.9 & 26.8 & 25.3 & 39 & 0.0 & 1.8 & $10.60^{+5.10}_{-3.30}$ \\
104 & 26.0 & 26.9 & 25.5 & 110 & 0.6 & 1.2 & $11.26^{+1.24}_{-2.66}$ \\
105 & 26.3 & 27.0 & 26.0 & 225 & 0.2 & 0.6 & $8.77^{+2.93}_{-0.17}$ \\
106 & 26.1 & 27.1 & 25.9 & 137 & 0.0 & 1.4 & $10.64^{+1.06}_{-2.54}$ \\
107 & 26.0 & 26.9 & 25.7 & 266 & 2.0 & 0.8 & $8.73^{+2.17}_{-0.13}$ \\
108 & 26.1 & 27.0 & 25.9 & 251 & 0.0 & 0.4 & $11.74^{+1.66}_{-3.64}$ \\
109 & 26.1 & 27.0 & 26.0 & 120 & 0.0 & 0.2 & $9.91^{+5.79}_{-0.91}$ \\
110 & 26.3 & 27.1 & 26.0 & 182 & 0.0 & 0.8 & $8.35^{+7.35}_{-0.25}$ \\
111 & 26.1 & 27.1 & 25.8 & 51 & 0.0 & 1.4 & $8.35^{+5.05}_{-0.65}$ \\
112 & 26.1 & 27.0 & 26.0 & 149 & 0.4 & 1.6 & $14.12^{+0.38}_{-6.42}$ \\
113 & 26.0 & 26.7 & 25.1 & 46 & 0.0 & 1.6 & $8.35^{+3.35}_{-0.65}$ \\
114 & 26.3 & 27.1 & 26.1 & 205 & 0.0 & 0.8 & $8.88^{+3.62}_{-0.28}$ \\
116 & 26.0 & 27.0 & 25.8 & 71 & 0.0 & 0.8 & $8.74^{+0.86}_{-0.14}$ \\
117 & 26.2 & 27.2 & 26.1 & 102 & 0.0 & 0.8 & $8.35^{+8.75}_{-0.25}$ \\
118 & 26.0 & 26.9 & 25.7 & 166 & 1.6 & 1.0 & $19.92^{+0.68}_{-9.02}$ \\
119 & 26.2 & 26.9 & 26.0 & 169 & 1.4 & 0.8 & $12.45^{+4.65}_{-0.75}$ \\
121 & 26.3 & 27.1 & 26.1 & 168 & 0.0 & 0.8 & $8.52^{+4.88}_{-0.42}$ \\
123 & 25.9 & 26.8 & 25.3 & 59 & 0.0 & 1.0 & $9.33^{+1.57}_{-0.73}$ \\
124 & 25.9 & 26.9 & 25.8 & 151 & 0.0 & 0.6 & $13.59^{+0.91}_{-4.59}$ \\
125 & 26.2 & 27.2 & 26.1 & 100 & 0.0 & 0.6 & $13.16^{+1.34}_{-2.26}$ \\
126 & 26.0 & 26.7 & 25.1 & 45 & 0.0 & 0.6 & $9.93^{+4.57}_{-0.93}$ \\
128 & 25.9 & 26.8 & 25.3 & 62 & 0.0 & 2.0 & $12.07^{+8.53}_{-4.77}$ \\
129 & 26.0 & 26.7 & 25.1 & 40 & 0.0 & 0.8 & $14.58^{+1.12}_{-5.58}$ \\
130 & 26.1 & 27.1 & 25.9 & 133 & 0.8 & 1.8 & $8.38^{+4.12}_{-1.08}$ \\
131 & 26.0 & 27.1 & 25.8 & 52 & 0.0 & 1.4 & $10.05^{+2.45}_{-2.35}$ \\
133 & 26.0 & 26.8 & 25.9 & 338 & 0.0 & 0.2 & $8.48^{+1.82}_{-0.38}$ \\
134 & 26.1 & 27.0 & 25.9 & 144 & 0.0 & 0.2 & $20.31^{+2.79}_{-4.61}$ \\
135 & 26.1 & 27.1 & 25.9 & 125 & 0.0 & 0.4 & $18.93^{+1.67}_{-9.33}$ \\
137 & 26.2 & 27.2 & 26.1 & 107 & 0.0 & 1.6 & $9.85^{+1.05}_{-2.15}$ \\
138 & 26.3 & 27.3 & 26.3 & 142 & 0.0 & 1.8 & $19.66^{+0.94}_{-11.96}$ \\
139 & 26.0 & 26.8 & 25.9 & 302 & 0.8 & 0.2 & $14.15^{+2.95}_{-0.75}$ \\
141 & 25.5 & 26.4 & 25.5 & 295 & 2.6 & 0.6 & $8.01^{+0.99}_{-0.31}$ \\
143 & 26.2 & 27.2 & 26.1 & 86 & 0.0 & 0.6 & $18.70^{+1.90}_{-6.20}$ \\
144 & 26.0 & 27.0 & 25.8 & 76 & 0.0 & 1.0 & $8.84^{+6.86}_{-0.24}$ \\
145 & 26.1 & 27.1 & 25.8 & 32 & 0.0 & 2.4 & $17.27^{+1.43}_{-9.97}$ \\
146 & 26.0 & 26.9 & 25.5 & 101 & 0.0 & 0.8 & $21.16^{+1.94}_{-8.66}$ \\
147 & 26.1 & 27.0 & 26.0 & 117 & 0.2 & 0.8 & $14.23^{+2.87}_{-3.93}$ \\
150 & 26.1 & 27.1 & 25.9 & 143 & 0.2 & 2.0 & $7.53^{+5.87}_{-0.23}$ \\
152 & 26.1 & 26.8 & 25.6 & 221 & 1.2 & 0.6 & $20.43^{+5.57}_{-7.03}$ \\
158 & 25.9 & 26.8 & 25.3 & 49 & 0.0 & 1.2 & $11.20^{+1.30}_{-2.60}$ \\
159 & 26.2 & 27.2 & 26.1 & 86 & 0.6 & 1.2 & $12.06^{+5.04}_{-4.36}$ \\
165 & 26.2 & 27.2 & 26.1 & 37 & 0.0 & 1.6 & $8.80^{+9.90}_{-1.10}$ \\
167 & 26.0 & 27.0 & 25.8 & 96 & 0.0 & 1.4 & $9.34^{+2.36}_{-2.04}$ \\
172 & 26.1 & 27.1 & 25.9 & 135 & 0.4 & 1.6 & $11.43^{+3.07}_{-4.13}$ \\
174 & 25.8 & 26.5 & 25.2 & 179 & 1.6 & 1.2 & $20.63^{+2.47}_{-12.93}$ \\
176 & 25.9 & 26.8 & 25.3 & 44 & 0.0 & 2.4 & $11.29^{+11.81}_{-3.99}$ \\
183 & 26.2 & 27.2 & 26.1 & 113 & 0.0 & 0.6 & $10.59^{+1.91}_{-0.29}$ \\
184 & 26.0 & 26.9 & 25.5 & 102 & 0.0 & 0.4 & $24.51^{+1.49}_{-14.21}$ \\
185 & 26.1 & 27.1 & 25.8 & 53 & 0.2 & 3.0 & $7.93^{+0.17}_{-0.23}$ \\
192 & 26.1 & 27.0 & 26.0 & 108 & 0.0 & 0.6 & $12.41^{+10.69}_{-0.71}$ \\
194 & 26.0 & 26.9 & 25.6 & 125 & 0.0 & 0.2 & $10.08^{+7.02}_{-0.48}$ \\
195 & 26.0 & 26.9 & 25.5 & 281 & 0.2 & 0.4 & $8.53^{+3.17}_{-0.43}$ \\
196 & 26.2 & 27.2 & 26.1 & 108 & 0.0 & 0.2 & $9.95^{+2.55}_{-0.35}$ \\
197 & 26.0 & 26.7 & 25.1 & 81 & 0.0 & 1.8 & $7.60^{+5.80}_{-0.30}$ \\
198 & 26.2 & 27.2 & 26.1 & 91 & 0.0 & 1.2 & $11.29^{+1.21}_{-3.19}$ \\
199 & 26.2 & 27.2 & 26.1 & 161 & 0.0 & 2.2 & $7.93^{+10.77}_{-0.63}$ \\
201 & 26.0 & 27.0 & 25.8 & 69 & 0.0 & 1.2 & $8.37^{+4.13}_{-0.67}$ \\
202 & 26.3 & 27.1 & 26.1 & 185 & 0.0 & 0.2 & $14.93^{+0.77}_{-4.03}$ \\
203 & 26.3 & 27.0 & 26.0 & 232 & 1.4 & 1.0 & $11.45^{+7.25}_{-2.85}$ \\
204 & 26.0 & 27.0 & 25.9 & 379 & 0.0 & 0.2 & $15.92^{+1.18}_{-5.62}$ \\
205 & 26.2 & 26.9 & 26.0 & 447 & 2.0 & 0.2 & $29.28^{+5.42}_{-13.58}$ \\
207 & 25.8 & 26.8 & 25.7 & 242 & 0.6 & 0.2 & $38.64^{+3.36}_{-19.94}$ \\
208 & 26.0 & 27.0 & 25.8 & 65 & 0.0 & 1.0 & $14.03^{+1.67}_{-5.43}$ \\
210 & 26.2 & 27.1 & 26.0 & 197 & 0.6 & 0.6 & $10.75^{+1.75}_{-1.15}$ \\
211 & 26.1 & 27.0 & 26.0 & 176 & 0.0 & 0.8 & $8.99^{+3.51}_{-0.39}$ \\
213 & 26.0 & 26.7 & 25.1 & 35 & 0.0 & 1.8 & $10.07^{+0.83}_{-2.37}$ \\
214 & 26.3 & 27.1 & 26.0 & 126 & 0.0 & 0.2 & $12.09^{+5.01}_{-1.19}$ \\
215 & 26.2 & 26.9 & 26.0 & 161 & 2.2 & 0.4 & $8.03^{+10.67}_{-0.33}$ \\
216 & 26.1 & 27.1 & 25.9 & 145 & 0.0 & 0.8 & $8.84^{+2.06}_{-0.24}$ \\
217 & 26.0 & 26.9 & 25.9 & 248 & 0.0 & 0.2 & $11.69^{+0.81}_{-3.09}$ \\
219 & 26.2 & 27.2 & 26.1 & 128 & 0.4 & 1.0 & $46.95^{+5.15}_{-38.35}$ \\
220 & 26.1 & 27.1 & 25.9 & 139 & 0.0 & 0.2 & $11.21^{+0.49}_{-1.61}$ \\
224 & 26.0 & 26.7 & 25.1 & 37 & 0.0 & 0.8 & $7.56^{+4.14}_{-0.26}$ \\
225 & 26.5 & 27.3 & 26.3 & 41 & 0.0 & 1.0 & $7.53^{+1.47}_{-0.23}$ \\
227 & 26.0 & 27.1 & 25.8 & 46 & 0.2 & 0.8 & $8.35^{+6.15}_{-0.25}$ \\
228 & 26.0 & 27.0 & 25.8 & 77 & 0.0 & 2.4 & $7.53^{+3.37}_{-0.23}$ \\
229 & 26.0 & 26.9 & 25.5 & 113 & 0.0 & 0.6 & $13.82^{+0.68}_{-4.22}$ \\
230 & 26.2 & 27.2 & 26.1 & 102 & 0.0 & 1.6 & $9.33^{+3.17}_{-1.63}$ \\
231 & 26.1 & 27.1 & 25.8 & 51 & 0.0 & 1.2 & $8.35^{+5.05}_{-0.25}$ \\
233 & 26.3 & 27.3 & 26.3 & 113 & 0.2 & 0.4 & $13.81^{+0.69}_{-2.91}$ \\
234 & 26.2 & 27.2 & 26.1 & 131 & 0.0 & 0.2 & $26.60^{+2.60}_{-15.70}$ \\
235 & 26.0 & 26.7 & 25.1 & 32 & 0.0 & 0.8 & $11.53^{+1.87}_{-2.93}$ \\
236 & 26.0 & 27.1 & 25.8 & 80 & 0.2 & 1.8 & $12.06^{+1.34}_{-4.76}$ \\
237 & 26.1 & 27.1 & 25.9 & 100 & 0.0 & 0.8 & $9.38^{+5.12}_{-0.38}$ \\
238 & 26.0 & 27.0 & 25.9 & 162 & 2.0 & 0.4 & $8.08^{+12.52}_{-0.38}$ \\
239 & 26.3 & 27.1 & 26.1 & 357 & 1.4 & 0.6 & $19.19^{+6.81}_{-10.19}$ \\
240 & 25.9 & 26.9 & 25.8 & 210 & 0.2 & 0.8 & $10.74^{+3.76}_{-1.14}$ \\
241 & 26.2 & 27.2 & 25.9 & 27 & 0.2 & 2.0 & $9.94^{+4.56}_{-2.64}$ \\
242 & 26.0 & 26.7 & 25.1 & 53 & 0.0 & 1.8 & $9.40^{+1.50}_{-2.10}$ \\
243 & 26.1 & 27.1 & 25.9 & 945 & 0.2 & 0.2 & $11.54^{+4.16}_{-0.64}$ \\
245 & 26.1 & 27.0 & 25.9 & 215 & 0.0 & 0.6 & $10.33^{+0.57}_{-1.73}$ \\
246 & 26.2 & 27.2 & 26.1 & 184 & 0.0 & 0.2 & $11.15^{+4.55}_{-2.15}$ \\
247 & 26.0 & 27.1 & 25.8 & 357 & 0.8 & 0.2 & $10.60^{+2.80}_{-1.00}$ \\
248 & 26.1 & 27.1 & 25.8 & 37 & 0.0 & 3.0 & $10.59^{+12.51}_{-3.29}$ \\
251 & 26.2 & 26.9 & 26.0 & 186 & 0.2 & 0.6 & $11.32^{+7.38}_{-1.02}$ \\
252 & 26.3 & 27.1 & 26.1 & 201 & 0.0 & 0.6 & $8.67^{+3.83}_{-0.57}$ \\
253 & 26.0 & 26.7 & 25.1 & 35 & 0.0 & 0.4 & $11.56^{+9.04}_{-1.96}$ \\
254 & 26.2 & 27.2 & 26.1 & 114 & 0.0 & 0.4 & $15.06^{+8.04}_{-5.46}$ \\
255 & 26.0 & 27.1 & 25.8 & 183 & 0.0 & 0.8 & $7.58^{+9.52}_{-0.28}$ \\
256 & 26.2 & 27.2 & 26.1 & 125 & 0.0 & 0.8 & $8.80^{+9.90}_{-0.20}$ \\
257 & 25.9 & 26.8 & 25.3 & 39 & 0.0 & 1.0 & $8.87^{+3.63}_{-0.27}$ \\
258 & 26.0 & 26.8 & 25.9 & 400 & 2.0 & 0.2 & $19.02^{+1.58}_{-9.42}$ \\
259 & 26.0 & 26.9 & 25.9 & 254 & 1.8 & 0.4 & $36.73^{+5.27}_{-25.83}$ \\
260 & 26.1 & 27.1 & 25.9 & 161 & 1.2 & 1.8 & $42.93^{+9.17}_{-35.63}$ \\
261 & 25.9 & 26.9 & 25.8 & 162 & 0.0 & 1.2 & $13.13^{+1.37}_{-4.53}$ \\
262 & 25.9 & 26.8 & 25.3 & 75 & 1.0 & 0.6 & $7.93^{+3.77}_{-0.23}$ \\
263 & 25.9 & 26.9 & 25.8 & 196 & 0.0 & 0.4 & $10.78^{+6.32}_{-0.48}$ \\
264 & 26.0 & 26.9 & 25.7 & 241 & 0.0 & 1.0 & $8.53^{+5.97}_{-0.43}$ \\
265 & 26.1 & 27.0 & 25.9 & 227 & 0.0 & 1.4 & $7.54^{+3.36}_{-0.24}$ \\
266 & 26.0 & 27.0 & 25.8 & 75 & 0.2 & 1.4 & $9.33^{+2.37}_{-2.03}$ \\
268 & 26.0 & 27.1 & 25.8 & 68 & 0.0 & 1.6 & $7.53^{+9.57}_{-0.23}$ \\
269 & 25.9 & 26.9 & 25.9 & 236 & 1.0 & 0.6 & $36.24^{+5.76}_{-23.74}$ \\
270 & 26.5 & 27.3 & 26.3 & 49 & 0.8 & 1.6 & $10.59^{+1.11}_{-2.89}$ \\
271 & 26.0 & 26.8 & 25.9 & 340 & 2.2 & 0.4 & $10.68^{+5.02}_{-0.38}$ \\
272 & 26.2 & 27.2 & 26.1 & 127 & 0.0 & 1.2 & $10.63^{+3.87}_{-2.53}$ \\
273 & 26.0 & 27.0 & 25.8 & 60 & 0.0 & 0.4 & $9.94^{+5.76}_{-0.94}$ \\
274 & 26.0 & 26.8 & 25.9 & 293 & 1.4 & 0.4 & $8.54^{+12.06}_{-0.44}$ \\
275 & 26.1 & 27.1 & 25.9 & 228 & 0.0 & 0.8 & $8.96^{+2.74}_{-0.36}$ \\
276 & 26.1 & 27.0 & 26.0 & 117 & 0.6 & 0.2 & $59.14^{+7.46}_{-45.74}$ \\
277 & 26.1 & 27.1 & 25.9 & 124 & 0.0 & 0.6 & $10.33^{+3.07}_{-1.73}$ \\
278 & 26.0 & 27.0 & 25.8 & 60 & 0.0 & 1.4 & $8.80^{+4.60}_{-0.20}$ \\
279 & 26.0 & 26.9 & 25.7 & 229 & 1.0 & 0.2 & $19.13^{+1.47}_{-7.43}$ \\
280 & 25.9 & 26.8 & 25.3 & 69 & 0.0 & 1.2 & $9.34^{+2.36}_{-1.64}$ \\
282 & 26.2 & 27.2 & 26.1 & 99 & 0.0 & 0.4 & $15.07^{+0.63}_{-2.57}$ \\
283 & 25.9 & 26.9 & 25.8 & 176 & 0.2 & 1.4 & $10.80^{+0.90}_{-1.80}$ \\
284 & 26.1 & 27.1 & 25.8 & 56 & 0.2 & 1.8 & $13.94^{+0.56}_{-6.64}$ \\
285 & 26.5 & 27.3 & 26.3 & 33 & 0.0 & 2.8 & $12.95^{+53.65}_{-5.65}$ \\
286 & 25.9 & 26.9 & 25.8 & 172 & 0.0 & 0.2 & $7.98^{+6.52}_{-0.28}$ \\
287 & 26.1 & 27.1 & 25.9 & 120 & 0.2 & 0.8 & $20.39^{+0.21}_{-11.39}$ \\
292 & 26.1 & 27.0 & 26.0 & 111 & 0.8 & 0.4 & $39.88^{+2.12}_{-26.48}$ \\
293 & 25.9 & 26.8 & 25.3 & 61 & 0.0 & 0.4 & $24.53^{+1.47}_{-14.93}$ \\
294 & 26.0 & 26.9 & 25.5 & 106 & 0.0 & 0.4 & $13.35^{+7.25}_{-2.45}$ \\
295 & 26.1 & 27.1 & 25.9 & 105 & 0.0 & 0.2 & $10.68^{+3.82}_{-0.38}$ \\
296 & 26.0 & 26.9 & 25.5 & 108 & 0.0 & 0.8 & $9.26^{+1.64}_{-0.66}$ \\
298 & 26.2 & 27.2 & 26.1 & 139 & 0.0 & 0.2 & $19.62^{+0.98}_{-10.02}$ \\
299 & 25.9 & 26.9 & 25.8 & 156 & 0.2 & 0.4 & $19.04^{+1.56}_{-7.34}$ \\
302 & 26.1 & 27.1 & 25.9 & 89 & 0.2 & 0.4 & $30.31^{+4.39}_{-16.91}$ \\
304 & 26.1 & 27.1 & 25.9 & 78 & 0.6 & 0.6 & $38.24^{+3.76}_{-26.54}$ \\
306 & 26.5 & 27.3 & 26.3 & 48 & 0.0 & 1.4 & $9.33^{+-9.33}_{-9.33}$ \\
SN1923A & 26.2 & 26.9 & 26.0 & 234 & 1.2 & 0.8 & $53.31^{+13.29}_{-43.71}$ \\
SN1950B & 26.1 & 27.1 & 25.8 & 39 & 0.0 & 3.2 & $7.56^{+3.34}_{-0.26}$ \\
SN1957D & 26.0 & 26.9 & 25.5 & 78 & 0.8 & 0.2 & $7.56^{+8.14}_{-0.26}$ \\
SN1983N & 25.5 & 26.4 & 25.5 & 293 & 0.2 & 0.2 & $20.43^{+5.57}_{-4.73}$ \\
     \enddata
\end{deluxetable*}

\clearpage
\begin{sidewaystable}
\vspace{7.0cm}
\begin{threeparttable}
\caption{Age Distribution Results from SNR 295\tablenotemark{a}}\label{tab:probDistro}
\centering
\begin{tabular}{llllllllllllll}
\hline
   (1) & (2) & (3) & (4) & (5) & (6) & (7) & (8) & (9) & (10) & (11) & (12) & (13) & (14)\\
   T1 &    T2 &  SFR (Best) &        -err &        +err &   PDF(Best) &   -err &   +err &   CDF(Best) &  CDF(16) & CDF(50) & CDF(84) &      M1 &    M2 \\
\hline
\hline
  4.0 &   4.5 &  0.0000e+00 &  0.0000e+00 &  6.5843e-04 &       0.000 &  0.000 &  0.075 &       0.000 &    0.000 &   0.000 &   0.033 &    52.1 &  66.6 \\
  4.5 &   5.0 &  0.0000e+00 &  0.0000e+00 &  5.7722e-04 &       0.000 &  0.000 &  0.074 &       0.000 &    0.000 &   0.018 &   0.057 &    42.0 &  52.1 \\
  5.0 &   5.6 &  0.0000e+00 &  0.0000e+00 &  6.3443e-04 &       0.000 &  0.000 &  0.089 &       0.000 &    0.002 &   0.034 &   0.083 &    34.7 &  42.0 \\
  5.6 &   6.3 &  0.0000e+00 &  0.0000e+00 &  5.4562e-04 &       0.000 &  0.000 &  0.087 &       0.000 &    0.012 &   0.049 &   0.108 &    29.2 &  34.7 \\
  6.3 &   7.1 &  0.0000e+00 &  0.0000e+00 &  5.1369e-04 &       0.000 &  0.000 &  0.091 &       0.000 &    0.023 &   0.065 &   0.133 &    26.0 &  29.2 \\
  7.1 &   7.9 &  0.0000e+00 &  0.0000e+00 &  4.3090e-04 &       0.000 &  0.000 &  0.086 &       0.000 &    0.033 &   0.080 &   0.156 &    23.1 &  26.0 \\
  7.9 &   8.9 &  0.0000e+00 &  0.0000e+00 &  4.4905e-04 &       0.000 &  0.000 &  0.099 &       0.000 &    0.044 &   0.097 &   0.182 &    20.6 &  23.1 \\
  8.9 &  10.0 &  0.0000e+00 &  0.0000e+00 &  4.8762e-04 &       0.000 &  0.000 &  0.118 &       0.000 &    0.057 &   0.117 &   0.213 &    18.7 &  20.6 \\
 10.0 &  11.2 &  1.5681e-03 &  1.1463e-03 &  2.4191e-04 &       0.242 &  0.192 &  0.249 &       0.242 &    0.155 &   0.265 &   0.412 &    17.1 &  18.7 \\
 11.2 &  12.6 &  0.0000e+00 &  0.0000e+00 &  3.6069e-04 &       0.000 &  0.000 &  0.111 &       0.242 &    0.172 &   0.284 &   0.440 &    15.7 &  17.1 \\
 12.6 &  14.1 &  0.0000e+00 &  0.0000e+00 &  3.7103e-04 &       0.000 &  0.000 &  0.126 &       0.242 &    0.190 &   0.305 &   0.473 &    14.5 &  15.7 \\
 14.1 &  15.8 &  0.0000e+00 &  0.0000e+00 &  3.5337e-04 &       0.000 &  0.000 &  0.134 &       0.242 &    0.210 &   0.327 &   0.507 &    13.4 &  14.5 \\
 15.8 &  17.8 &  0.0000e+00 &  0.0000e+00 &  3.5393e-04 &       0.000 &  0.000 &  0.148 &       0.242 &    0.230 &   0.351 &   0.544 &    12.5 &  13.4 \\
 17.8 &  20.0 &  0.0000e+00 &  0.0000e+00 &  3.8747e-04 &       0.000 &  0.000 &  0.175 &       0.242 &    0.254 &   0.380 &   0.585 &    11.7 &  12.5 \\
 20.0 &  22.4 &  3.2132e-35 &  3.2132e-35 &  3.5257e-04 &       0.000 &  0.000 &  0.179 &       0.242 &    0.279 &   0.409 &   0.629 &    10.9 &  11.7 \\
 22.4 &  25.1 &  1.9962e-03 &  1.3454e-03 &  4.9830e-05 &       0.688 &  0.469 &  0.165 &       0.930 &    0.579 &   0.755 &   0.892 &    10.3 &  10.9 \\
 25.1 &  28.2 &  0.0000e+00 &  0.0000e+00 &  2.7900e-04 &       0.000 &  0.000 &  0.178 &       0.930 &    0.616 &   0.788 &   0.919 &     9.6 &  10.3 \\
 28.2 &  31.6 &  1.6137e-04 &  1.6137e-04 &  6.6697e-04 &       0.070 &  0.070 &  0.383 &       1.000 &    0.741 &   0.874 &   0.978 &     9.0 &   9.6 \\
 31.6 &  35.5 &  0.0000e+00 &  0.0000e+00 &  2.8913e-04 &       0.000 &  0.000 &  0.220 &       1.000 &    0.793 &   0.913 &   1.000 &     8.6 &   9.0 \\
 35.5 &  39.8 &  0.0000e+00 &  0.0000e+00 &  2.4359e-04 &       0.000 &  0.000 &  0.211 &       1.000 &    0.846 &   0.955 &   1.000 &     8.1 &   8.6 \\
 39.8 &  44.7 &  0.0000e+00 &  0.0000e+00 &  2.2500e-04 &       0.000 &  0.000 &  0.217 &       1.000 &    0.909 &   1.000 &   1.000 &     7.7 &   8.1 \\
 44.7 &  50.1 &  0.0000e+00 &  0.0000e+00 &  2.0419e-04 &       0.000 &  0.000 &  0.220 &       1.000 &    1.000 &   1.000 &   1.000 &     7.3 &   7.7 \\
\hline
\end{tabular}
\begin{tablenotes}
\centering
\item[a]Similar tables are available for all fitted SNRs with inferred progenitor masses in the supplemental materials.
\end{tablenotes}
\end{threeparttable}
\end{sidewaystable}

\clearpage 

\end{document}